\documentclass[twocolumn,english]{revtex4-1}
\usepackage{mathptmx}
\usepackage{lmodern}
\usepackage[T1]{fontenc}
\usepackage[latin9]{inputenc}
\usepackage{amsmath}
\usepackage{amssymb}
\usepackage{graphicx}
\usepackage{esint}

\makeatletter

\providecommand{\tabularnewline}{\\}

\@ifundefined{textcolor}{}
{%
 \definecolor{BLACK}{gray}{0}
 \definecolor{WHITE}{gray}{1}
 \definecolor{RED}{rgb}{1,0,0}
 \definecolor{GREEN}{rgb}{0,1,0}
 \definecolor{BLUE}{rgb}{0,0,1}
 \definecolor{CYAN}{cmyk}{1,0,0,0}
 \definecolor{MAGENTA}{cmyk}{0,1,0,0}
 \definecolor{YELLOW}{cmyk}{0,0,1,0}
}

\usepackage{babel}

\makeatother

\usepackage{babel}
\begin{document}

\title{Real-Space Renormalization Group for Spectral Properties of Hierarchical
Networks}

\author{Stefan Boettcher and Shanshan Li}

\affiliation{Dept. of Physics, Emory University, Atlanta, GA 30322; USA}

\homepage{http://www.physics.emory.edu/faculty/boettcher/}

\begin{abstract}
We derive the determinant of the Laplacian for the Hanoi networks
and use it to determine their number of spanning trees (or graph complexity)
asymptotically. While spanning trees generally proliferate with increasing
average degree, the results show that modifications within the basic
patterns of design of these hierarchical networks can lead to significant
variations in their complexity. To this end, we develop renormalization
group methods to obtain recursion equations from which many spectral
properties can be obtained. This provides the basis for future applications
to explore the physics of several dynamic processes. 
\end{abstract}
\maketitle

\section{Introduction\label{sec:Introduction}}

Understanding the dynamics of hierarchical systems is rapidly becoming
a subject of wide interest within the study of complex networks~\cite{Trusina04,Clauset08,Fischer08,SWPRL,Hasegawa13b}.
This may seem surprising, as hierarchies are not a new subject in
complexity~\cite{Simon62}, and considering that the ultimate hierarchical
system -- a tree -- has long been studied in many variations and for
many types of dynamics because exact results can be obtained. Known
as Bethe approximation in statistical physics~\cite{Pathria,Mezard06},
trees locally resemble sparse mean-field models in physics, due to
their infinite dimensionality. Beyond the mean-field statistical models,
however, there exists a much richer dynamics arises from peculiar
structural features of hierarchical networks such as high degree of
clustering and modularity. In many realistic situations, such hierarchies
are embedded in some finite-dimensional space~\cite{Dorogovtsev08,barthelemy_spatial_2010},
which can lead to an entirely new set of phenomena. Examples of such
embedded hierarchical systems contain transport and control systems,
they have been observed in the brain~\cite{Moretti13,Meunier09},
or they have been studied for their novel synthetic critical behavior~\cite{Hinczewski06,Boettcher09c,Boettcher10c,BoBr12,PhysRevLett.108.255703},
such as explosive transitions that are purely induced by the geometry
in percolation~\cite{Boettcher11d} or synchronization~\cite{Shanshan15}.
Recently, hierarchical networks based on Dyson's model have shown
to allow for the existence of a number of metastable states, which
can be used to study the modular architecture and parallel processing
in neuron networks as well as the ergodicity breakdown for the stochastic
process\ \cite{agliari15retrieval,agliari15topological,agliari2015hierarchical}. 

Key to understanding the mechanisms at the core of those novel phenomena
lies within the geometry of these hierarchies. These geometric properties
are inevitably tied to spectral properties of their network Laplacians~\cite{Biggs74}.
Here, we investigate a recently proposed class of hierarchical networks~\cite{SWPRL}
within the simplest of spatial embeddings -- a simple line -- for
which we can obtain many spectral properties exactly using the renormalization
group (RG)~\cite{Pathria}.For these Hanoi networks, we study some
basic properties of their Laplacians. In particular, we derive sets
of recursion relations that allow to study their secular equations
(also called characteristic polynomials), whose zeros provide all
eigenvalues, to arbitrary accuracy. We determine the asymptotic scaling
of their determinants with system size. These determinants can be
used as a generator for many other asymptotic properties.

The Laplacian matrix is given by 
\begin{eqnarray}
\mathbf{L}_{i,j} & = & d_{i}\delta_{i,j}-A_{i,j},\label{eq:Laplacian}
\end{eqnarray}
where $d_{i}$ specifies the degree of the $i$-th site and $A_{i,j}$
is the adjacency matrix of the network. Since we assume that the links
in the networks are undirected, ${\bf A}$, and hence ${\bf L}$,
are symmetric. We further assume that there are no external links,
which implies the vanishing of all row or column sums in ${\bf L}$,
$\sum_{i}\mathbf{L}_{i,j}=\sum_{j}\mathbf{L}_{i,j}=0$. The fundamental
property characterizing the Laplacian matrix is its spectrum of eigenvalues,
the solutions $\lambda_{i}$ of the secular equation 
\begin{eqnarray}
\det\left[{\bf L}-\lambda{\bf 1}\right] & = & 0.\label{eq:secular}
\end{eqnarray}
This spectrum is highly nontrivial for the Hanoi networks, and we
will not be able to describe it in much detail here. But we can provide
an RG approach that reduces the effort \emph{exponentially} from solving
$2^{k}\times2^{k}$ determinants to $k$ iterations of a closed set
of RG recursion equations for any desired quantity, where $N=2^{k}$
refers to the number of sites in the network. Numerous aspects can
be extracted in closed form asymptotically. 

The spectrum of network Laplacians features in many practical applications.
The scaling of the ratio between largest and smallest eigenvalue indicates
the synchronizability of a network~\cite{Barahona02}, which also
can be approximated by a sum over all eigenvalues (inverted) when
related to a random deposition process~\cite{Korniss03,SITIS12}.
Permeability and well-connectedness of networks can be defined in
terms of their smallest eigenvalue\ \cite{maas87permeability}. The
spectrum further features prominently in quantum transport~\cite{Boettcher11b},the
behavior of continuous-time quantum search algorithms~\cite{Childs04,PhysRevA.82.012305},
graph partitioning\ \cite{pothen90partitioning,HL} and image processing\ \cite{peinecke07image},
just to name a few examples. Of course, Laplacian spectra also determine
the characteristic frequencies of mechanical vibrations, from which
connectivity between interacting units within the membrane can be
identified\ \cite{fisher66shape}. Thus, there has long been strong
motivation to study such spectra particularly on fractal networks,
where its properties can be explored in great detail~\cite{PhysRevB.28.3110,Rammal84,fukushima92spectral,rammal83random,shima91eigenvalue,teplyaev00gradients}.

\begin{figure}
\begin{centering}
\includegraphics[bb=8bp 8bp 540bp 699bp,clip,angle=270,width=0.8\columnwidth]{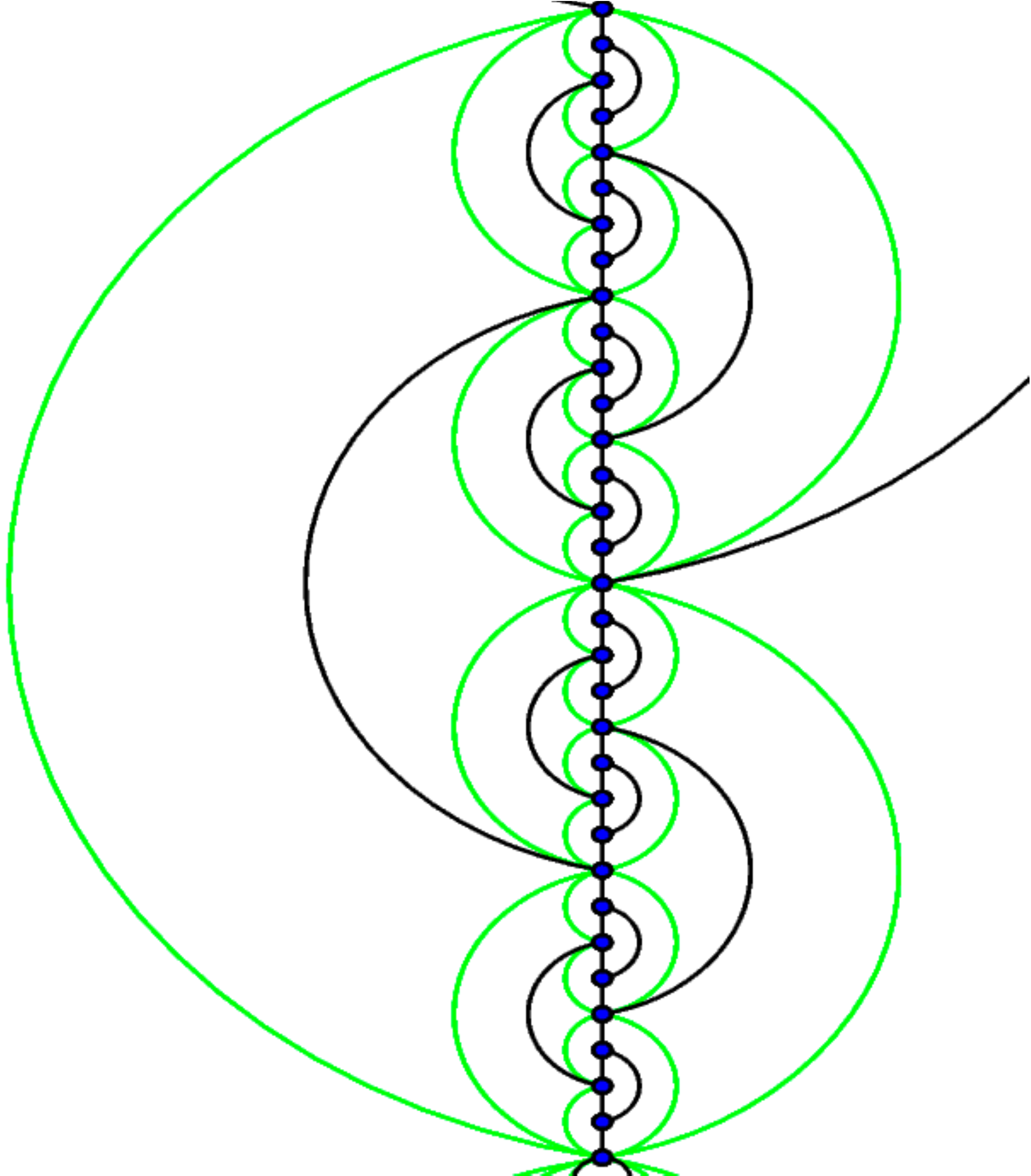} 
\par\end{centering}
\protect\protect\protect\caption{\label{fig:HN3}Depiction of the Hanoi networks HN3 and HN5. The 3-regular
network HN3 corresponds to the solid (black) lines alone, while HN5
in addition also consists of the (green-) shaded lines. For HN5, sites
on the lowest level of the hierarchy have degree 3, then degree 5,
7, etc, concerning a fraction of 1/2, 1/4, 1/8, etc., of all sites,
which makes for an average degree 5 in this network. Note that both
networks are planar.}
\end{figure}

\begin{figure}[t!]
\begin{centering}
\includegraphics[bb=228bp 559bp 570bp 1122bp,clip,angle=270,width=0.8\columnwidth]{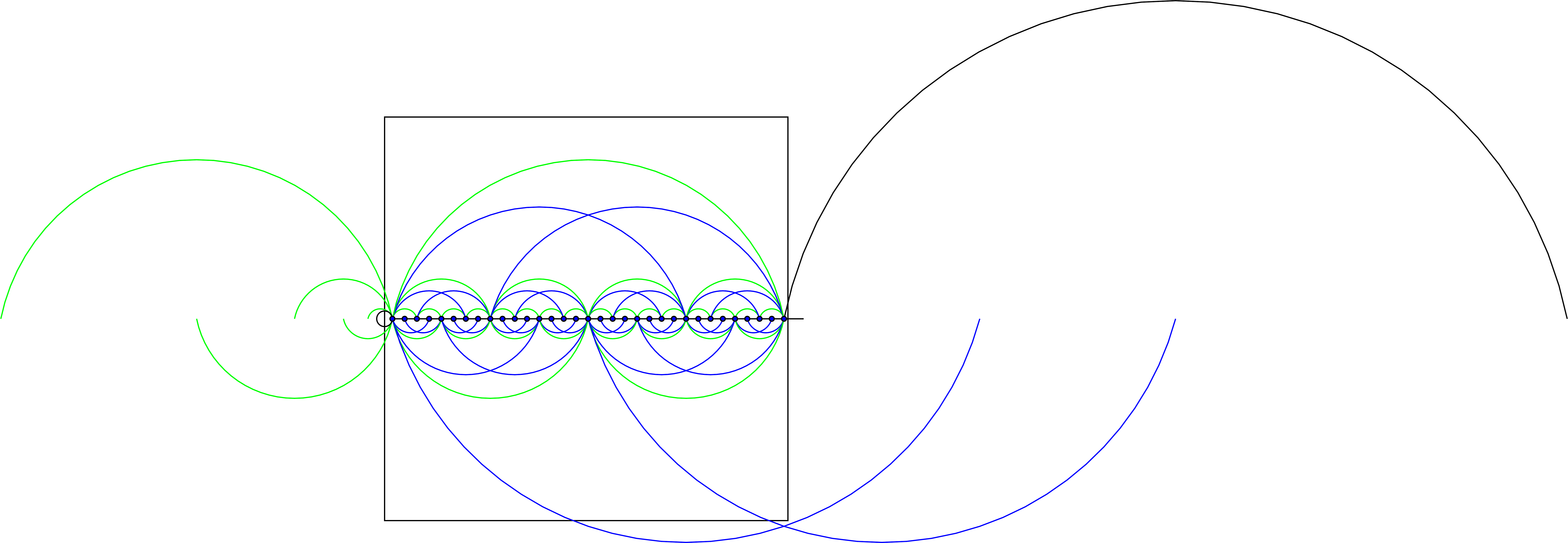} 
\par\end{centering}
\protect\protect\protect\caption{Depiction of the nonplanar Hanoi networks HNNP and HN6. Again, starting
from a 1d-backbone (black lines), in HNNP a set of long-range links
(blues-shades lines) is added that break planarity but maintain the
hierarchical pattern set out in Eq.~(\ref{eq:numbering}) that leads
to a network of average degree 4. If we add again the same links that
distinguished HN3 and HN5 in Fig.~\ref{fig:HN3}, we obtain HN6 with
average degree 6. In all cases, the RG on these networks remains exact.\label{fig:HN-NP}}
\end{figure}

Here, we will only focus on the simplest case of the scaling of the
determinant itself, obtained by taking $\lambda\to0$ in Eq.~(\ref{eq:secular}),
which provides the number of spanning trees, 
\begin{equation}
\#_{{\rm ST}}=-\lim_{\lambda\to0}\frac{\det\left[{\bf L}-\lambda{\bf 1}\right]}{\lambda N},
\label{eq:nST}
\end{equation}
sometimes also referred to as ``graph complexity''~\cite{Lyons05}.
Eq. (\ref{eq:nST}) is one of the oldest results in algebraic graph
theory, due to Kirchhoff (1847)~\cite{Biggs74}. Spanning trees describe
the size of the attractor state in the self-organized critical sandpile
model~\cite{Dhar99}, they characterize the optimal paths between
any two sites in a network~\cite{wutransport:06}, and are also related
to optimal synchronizability of a network~\cite{Nishikawa:06}. The
number of spanning trees is of fundamental interest in mathematics
and physics. For example, it is related to the partition function
of $q$-state Potts model in the limit $q\rightarrow0$~\cite{Dhar77,Wu82}
Thus, studies on the asymptotic growth of spanning trees are well
motivated not only for regular lattices\ \cite{shrock:00}, but also
on self-similar structures~\cite{Chang07,teufl06spanning,teufl11Strees}.
On these networks, the number of spanning trees $\#_{ST}$ exponentially
increases with $N$, which can be characterized by the \emph{tree
entropy}, which is the entropy-density of spanning trees~\cite{Lyons05,Chang07,Teufl10}, 
\begin{eqnarray}
s & = & \lim_{N\rightarrow\infty}\frac{\ln\left(\#_{{\rm ST}}\right)}{N}.
\label{eq:STentropy}
\end{eqnarray}
The number of spanning trees on Hanoi networks can be easily derived
from our RG procedure, allowing us to explore the role of geometric
structures on its asymptotic growth. 

\section{Structure of Hierarchical Networks\label{sec:Network-Structure}}

The Hanoi networks we discuss in this paper were introduced first
in Ref. \cite{SWPRL}. Each of the networks considered in this paper
possesses a simple geometric backbone, a one-dimensional line of sites
$N$. in which each site is at least connected to its nearest neighbor
left and right on the backbone. To generate the small-world hierarchy
in these networks, consider parameterizing any number $n>0$ \emph{uniquely}
in terms of two other integers $(i,j)$, $i\geqslant1$ and $1\leqslant j\leqslant2^{k-i}$,
via 
\begin{eqnarray}
n & = & 2^{i-1}\left(2j-1\right),\label{eq:numbering}
\end{eqnarray}
which motivated the name of the networks %
\footnote{The hierarchy index $i$ of sites $n$ in Eq. (\ref{eq:numbering})
resembles the sequence by which discs are moved in the famous ``Tower
of Hanoi'' problem. Unfortunately, there also exists a ``Hanoi graph''
(see Weisstein, Eric W. \textquotedbl{}Hanoi Graph.\textquotedbl{}
From MathWorld--A Wolfram Web Resource. http://mathworld.wolfram.com/HanoiGraph.html),
essentially the dual of the Sierpinski gasket, that should not be
confused with our networks here.%
}. 

The networks can be considered either closed into a loop of $N=2^{k}$
sites or in a linear arrangement with $N=2^{k}+1$ sites, a difference
that typically does not effect the RG and any leading asymptotic property.
Their recursive pattern is far more easily to illustrate when drawn
linearly, as for HN3 and HN5 in Fig. \ref{fig:HN3} and for HNNP and
HN6 in Fig. \ref{fig:HN-NP}, although in our calculations we avoid
spurious edge-effects by considering periodic loops, where sites $n=0$
and $n=N=2^{k}$ are \emph{identical} to each other. For convenient
comparisons, we call the ordinary one-dimensional loop HN2 (for Hanoi
Network of degree 2). The details of the design and most of their
geometric properties have been discussed at length elsewhere \cite{Boettcher10c}.

\section{Obtaining the RG-Recursions\label{sec:Obtaining-the-RG-Recursions}}

In this Section, we show how to obtain the recursion equations for
the RG-flow of our networks that describes the asymptotic properties
of the lattice Laplacian. To that end, we employ the well-known formal
identity~\cite{Ramon97}, 
\begin{widetext}
\begin{eqnarray}
\frac{1}{\sqrt{\det L_{k}^{(\alpha)}}} & = & \dotsintop_{-\infty}^{\infty}\left(\prod_{i=n}^{N}\frac{dx_{n}}{\sqrt{\pi}}\right)\exp\left\{ -\sum_{n=1}^{N}\sum_{m=1}^{N}x_{n}\left(L_{k}^{(\alpha)}\right)_{n,m}x_{m}\right\} ,\label{eq:gaussint}
\end{eqnarray}
\end{widetext}
for the evaluation of the determinant of $L_{k}^{(\alpha)}$ for each
network HN$\alpha$ with $\alpha\in\left\{ 2,3,5,NP,6\right\} $ of
size $N=2^{k}$. (Since Laplacians are singular matrices, formally,
some integrals will not converge and need regulation, however, these
singularities will be dealt with explicitly below in the final step
of the RG.) We can group our networks into two classes of equivalent
topologies, as respectively Figs.~\ref{fig:HN3} and \ref{fig:HN-NP}
suggest: HN3 is contained within HN5 and HNNP is contained within
HN6, while the simple loop (HN2) is contained within both, of course.
In addition, removing the backbone of HNNP decomposes it into three
separate, loop-less tree structures that have been studied in Ref.~\cite{Hasegawa13c}.

The following calculations are purely formal, as some of the integrals
may not converge in Eq.~(\ref{eq:gaussint}), in which case we would
have to insert some form of regulation. Yet, the integrals merely
serve as generators for the desired RG-recursions, essentially. Alternative
algebraic means to evaluate the determinants that do not involve integrals
have been used for a specific application previously~\cite{PhysRevB.28.3110,Boettcher11b}.
However, the approach taken here provides entirely equivalent results
with a clearer perspective on the topological transformations that
are involved. Furthermore, the current approach is closely linked
to the description of a path-integral for a free field-theory~\cite{Ramon97}
on a given network, which affords certain extensions of the methods,
for example, by inserting source-terms $\sum_{n=1}^{N}J_{n}x_{n}$
into the exponential in Eq.~(\ref{eq:gaussint}) as generators for
more complicated observables. When these sources are either uniform
($J_{n}\equiv J$ f.a. $n$), localized ($J_{n}=J\delta_{n,x}$),
or hierarchically staggered {[}$J_{n}=J_{i(n)}$ according to Eq.~(\ref{eq:numbering}){]},
exact RG-recursions may still be obtained.

\subsection{Renormalization Group calculation for the determinant of HN3 and
HN5\label{sub:RG_HN3}}

Here, we evaluate the most general $2^{k}\times2^{k}$-determinant
for the matrix $L_{k}^{(\alpha)}$ that is used in the analysis of
HN3 and HN5 (and HN2) below. The properties of each determinant emerges
via renormalization from $L_{k}^{(\alpha)}$ for a different set of
``bare'' parameters, while the RG-recursions themselves remain the
same for each $\alpha\in\left\{ 2,3,5\right\} $. These parameters
describing the renormalized weights of effective links between sites,
while their bare values serve as initial conditions on these RG recursion
equations. We define $L_{k}^{(2)}$ as the matrix of a one-dimensional
loop of $2^{k}$ sites (which we may call HN2). $L_{k}^{(3)}$ and
$L_{k}^{(5)}$ are respectively the matrices for the HN3 and HN5 networks.
Instead of providing a formal description of $L_{k}^{(\alpha)}$ for
general size $2^{k}$, we simply illustrate its generic recursive
pattern for the case $k=4$: 
\begin{widetext}
\begin{eqnarray}
L_{k=4}^{(3,5)} & = & \left[\begin{array}{cccccccccccccccc}
q_{4} & -p_{0} & -l_{1} & 0 & -l_{2} & 0 & 0 & 0 & -2l_{3} & 0 & 0 & 0 & -l_{2} & 0 & -l_{1} & -p_{0}\\
-p_{0} & q_{1} & -p_{0} & -p_{1} & 0 & 0 & 0 & 0 & 0 & 0 & 0 & 0 & 0 & 0 & 0 & 0\\
-l_{1} & -p_{0} & q_{2} & -p_{0} & -l_{1} & 0 & -p_{2} & 0 & 0 & 0 & 0 & 0 & 0 & 0 & 0 & 0\\
0 & -p_{1} & -p_{0} & q_{1} & -p_{0} & 0 & 0 & 0 & 0 & 0 & 0 & 0 & 0 & 0 & 0 & 0\\
-l_{2} & 0 & -l_{1} & -p_{0} & q_{3} & -p_{0} & -l_{1} & 0 & -l_{2} & 0 & 0 & 0 & -p_{3} & 0 & 0 & 0\\
0 & 0 & 0 & 0 & -p_{0} & q_{1} & -p_{0} & -p_{1} & 0 & 0 & 0 & 0 & 0 & 0 & 0 & 0\\
0 & 0 & -p_{2} & 0 & -l_{1} & -p_{0} & q_{2} & -p_{0} & -l_{1} & 0 & 0 & 0 & 0 & 0 & 0 & 0\\
0 & 0 & 0 & 0 & 0 & -p_{1} & -p_{0} & q_{1} & -p_{0} & 0 & 0 & 0 & 0 & 0 & 0 & 0\\
-2l_{3} & 0 & 0 & 0 & -l_{2} & 0 & -l_{1} & -p_{0} & q_{4} & -p_{0} & -l_{1} & 0 & -l_{2} & 0 & 0 & 0\\
0 & 0 & 0 & 0 & 0 & 0 & 0 & 0 & -p_{0} & q_{1} & -p_{0} & -p_{1} & 0 & 0 & 0 & 0\\
0 & 0 & 0 & 0 & 0 & 0 & 0 & 0 & -l_{1} & -p_{0} & q_{2} & -p_{0} & -l_{1} & 0 & -p_{2} & 0\\
0 & 0 & 0 & 0 & 0 & 0 & 0 & 0 & 0 & -p_{1} & -p_{0} & q_{1} & -p_{0} & 0 & 0 & 0\\
-l_{2} & 0 & 0 & 0 & -p_{3} & 0 & 0 & 0 & -l_{2} & 0 & -l_{1} & -p_{0} & q_{3} & -p_{0} & -l_{1} & 0\\
0 & 0 & 0 & 0 & 0 & 0 & 0 & 0 & 0 & 0 & 0 & 0 & -p_{0} & q_{1} & -p_{0} & -p_{1}\\
-l_{1} & 0 & 0 & 0 & 0 & 0 & 0 & 0 & 0 & 0 & -p_{2} & 0 & -l_{1} & -p_{0} & q_{2} & -p_{0}\\
-p_{0} & 0 & 0 & 0 & 0 & 0 & 0 & 0 & 0 & 0 & 0 & 0 & 0 & -p_{1} & -p_{0} & q_{1}
\end{array}\right].\label{eq:H_k^a}
\end{eqnarray}
\end{widetext}

The bare parameters $q_{i}$ on the diagonal refer to on-site properties
of each site $n$ that belongs to the $i$-th hierarchy as determined
by Eq.~(\ref{eq:numbering}). For the case of the lattice Laplacian
here, for example, $q_{i}$ is simply the degree of that site. The
off-diagonal parameters label extant or potentially emerging links
between sites that are undirected, so the matrix is symmetric. To
keep parameters fundamentally non-negative, we insert negative signs
explicitly, so that the bare values for all $p_{i}$ is uniformly
unity, however, they renormalize differently for each $i$. (Here,
level $i=0$ refers to nearest-neighbor links in the backbone.) While
the $p_{i}$ correspond to the solid black lines in Fig.~\ref{fig:HN3},
the parameters $l_{i}$ refer to those long-range links shaded in
green in Fig.~\ref{fig:HN3}, whose addition make up HN5. (Entries
like $-2l_{k-1}$ near the highest level of the hierarchy correspond
to a convenient choice in imposing periodic boundary conditions on
the network.) Although all $l_{i}$ are originally zero for HN3, we
still need to consider them, as they \emph{emerge} as a relevant parameter~\cite{Pathria}
during the RG, even for HN3. If $p_{i}=0$ for all $i>0$, Eq.~(\ref{eq:H_k^a})
reduces to the tridiagonal matrix for a simple loop we call HN2. Note
that we have imposed periodic boundary conditions by identifying site
$N=2^{k}$ with site $n=0$, where the site indices run from $n=0$
to $n=2^{k}-1$.

Employing the binary decomposition of the integer labels implied by
Eq.~(\ref{eq:numbering}) for the sites on the network backbone,
we can write Eq.~(\ref{eq:gaussint}) as 
\begin{widetext}
\begin{eqnarray}
\left[\det L_{k}^{(3,5)}\right]^{-\frac{1}{2}} & = & I_{k}\iintop_{-\infty}^{\infty}\frac{dx_{0}dx_{2^{k-1}}}{\pi}\exp\left\{ -q_{k}\left(x_{0}^{2}+x_{2^{k-1}}^{2}\right)+4l_{k-1}x_{0}x_{2^{k-1}}\right\}  \\
 &  & \dotsintop_{-\infty}^{\infty}\left(\prod_{i=1}^{k-1}\prod_{j=1}^{2^{k-i}}\frac{dx_{2^{i-1}\left(2j-1\right)}}{\sqrt{\pi}}\right)\label{eq:HN3int}\quad\exp\left\{ -\sum_{i=1}^{k-1}q_{i}\sum_{j=1}^{2^{k-i}}x_{2^{i-1}\left(2j-1\right)}^{2}+2\sum_{i=1}^{k-2}l_{i}\sum_{j=1}^{2^{k-i}}x_{2^{i-1}\left(2j-2\right)}x_{2^{i-1}\left(2j\right)}\right.\nonumber \\
 &  & \qquad\qquad\qquad\qquad\qquad\qquad\left.+2\sum_{i=1}^{k-1}p_{i}\sum_{j=1}^{2^{k-i-1}}x_{2^{i-1}\left(4j-3\right)}x_{2^{i-1}\left(4j-1\right)}+2p_{0}\sum_{j=1}^{2^{k-1}}x_{2j-1}\left(x_{2j-2}+x_{2j}\right)\right\} .\nonumber 
\end{eqnarray}
\end{widetext}
The factor $I_{k}$, initially unity, captures the contribution of
integrals from any prior RG-step.

To solve $\det L_{k}^{(3,5)}$ recursively, we integrate only over
all variables $x$ of \emph{odd} index {[}those with $i=1$ in Eq.
(\ref{eq:HN3int}){]}. To that end, we focus on the case $i=1$ in
the product of integrals in Eq.~(\ref{eq:HN3int}) and re-write $\left.\prod_{j=1}^{2^{k-i}}dx_{2^{i-1}\left(2j-1\right)}\right|_{i=1}=\prod_{j=1}^{2^{k-1}}dx_{2j-1}=\prod_{j=1}^{2^{k-2}}dx_{4j-3}dx_{4j-1}$ to get: 
\begin{widetext}
\begin{eqnarray}
 &  & \prod_{j=1}^{2^{k-2}}\iintop_{-\infty}^{\infty}\frac{dx_{4j-3}dx_{4j-1}}{\pi}\exp\left\{ -q_{1}\left(x_{4j-3}^{2}+x_{4j-1}^{2}\right)+2p_{1}x_{4j-3}x_{4j-1}\right.\left.+2p_{0}\left[x_{4j-3}\left(x_{4j-4}+x_{4j-2}\right)+x_{4j-1}\left(x_{4j-2}+x_{4j}\right)\right]\right\} ,\nonumber \\
 & = & \prod_{j=1}^{2^{k-2}}\left(q_{1}^{2}-p_{1}^{2}\right)^{-\frac{1}{2}}\exp\left\{ \frac{q_{1}p_{0}^{2}}{q_{1}^{2}-p_{1}^{2}}\left[\left(x_{4j-4}+x_{4j-2}\right)^{2}+\left(x_{4j-2}+x_{4j}\right)^{2}\right]\right.\left.+\frac{2p_{1}p_{0}^{2}}{q_{1}^{2}-p_{1}^{2}}\left(x_{4j-4}+x_{4j-2}\right)\left(x_{4j-2}+x_{4j}\right)\right\} ,\\
 & = & \left(q_{1}^{2}-p_{1}^{2}\right)^{-2^{k-3}}\exp\left\{ \frac{2q_{1}p_{0}^{2}}{q_{1}^{2}-p_{1}^{2}}\left(x_{0}^{2}+x_{2^{k-1}}^{2}\right)+\frac{2q_{1}p_{0}^{2}}{q_{1}^{2}-p_{1}^{2}}\sum_{i=2}^{k-2}\sum_{j=1}^{2^{k-i-1}}x_{2^{i-1}\left(2j-1\right)2}^{2}\right.\nonumber \\
 &  & \qquad\qquad\qquad\qquad\qquad\qquad\left.+\frac{2p_{0}^{2}}{q_{1}-p_{1}}\sum_{j=1}^{2^{k-2}}x_{(2j-1)2}^{2}+\frac{2p_{0}^{2}}{q_{1}-p_{1}}\sum_{j=1}^{2^{k-1}}x_{2j-2}x_{2j}+\frac{2p_{1}p_{0}^{2}}{q_{1}^{2}-p_{1}^{2}}\sum_{j=1}^{2^{k-2}}x_{(2j-2)2}x_{(2j)2}\right\} .\nonumber 
\end{eqnarray}
\end{widetext}
With that result, the remaining integral over the even-indexed variables
can be written as 
\begin{widetext}
\begin{eqnarray}
\left[\det L_{k}^{(3,5)}\right]^{-\frac{1}{2}} & = & I_{k}\left(q_{1}^{2}-p_{1}^{2}\right)^{-2^{k-3}}\iintop_{-\infty}^{\infty}\frac{dx_{0}dx_{2^{k-1}}}{\pi}\exp\left\{ -\left[q_{k}-\frac{2q_{1}p_{0}^{2}}{q_{1}^{2}-p_{1}^{2}}\right]\left(x_{0}^{2}+x_{2^{k-1}}^{2}\right)+4l_{k-1}x_{0}x_{2^{k-1}}\right\} \nonumber \\
 &  & \dotsintop_{-\infty}^{\infty}\left(\prod_{i=1}^{k-2}\,\prod_{j=1}^{2^{k-1-i}}\frac{dx_{2^{i}\left(2j-1\right)}}{\sqrt{\pi}}\right)\exp\left\{ -\sum_{i=2}^{k-2}\left[q_{i+1}-\frac{2q_{1}p_{0}^{2}}{q_{1}^{2}-p_{1}^{2}}\right]\sum_{j=1}^{2^{k-1-i}}x_{2^{i}\left(2j-1\right)}^{2}\right.\nonumber \\
 &  & \qquad\qquad\qquad-\left[q_{2}-\frac{2p_{0}^{2}}{q_{1}-p_{1}}\right]\sum_{j=1}^{2^{k-2}}x_{2\left(2j-1\right)}^{2}+2\sum_{i=1}^{k-2}p_{i+1}\sum_{j=1}^{2^{k-i-2}}x_{2^{i}\left(4j-3\right)}x_{2^{i}\left(4j-1\right)}\label{eq:HN3int_reno}\\
 &  & \qquad\qquad\qquad+2\sum_{i=2}^{k-3}l_{i+1}\sum_{j=1}^{2^{k-i-1}}x_{2^{i}\left(2j-2\right)}x_{2^{i}\left(2j\right)}+2\left(l_{2}+\frac{p_{1}p_{0}^{2}}{q_{1}^{2}-p_{1}^{2}}\right)\sum_{j=1}^{2^{k-2}}x_{2(2j-2)}x_{2(2j)}\nonumber \\
 &  & \qquad\qquad\qquad\left.+2\left(l_{1}+\frac{p_{0}^{2}}{q_{1}-p_{1}}\right)\sum_{j=1}^{2^{k-2}}x_{2(2j-1)}\left(x_{2(2j-2)}+x_{2(2j)}\right)\right\} .\nonumber 
\end{eqnarray}
\end{widetext}
Substituting $x_{i}^{\prime}=x_{2i}$, this expression is \emph{identical}
in form with Eq.~(\ref{eq:HN3int}), and we can identify 
\begin{eqnarray}
q{}_{1}^{\prime} & = & q_{2}-2\frac{p_{0}^{2}}{q_{1}-p_{1}},\nonumber \\
q{}_{i}^{\prime} & = & q_{i+1}-2\frac{q_{1}p_{0}^{2}}{q_{1}^{2}-p_{1}^{2}},\qquad(i\geq2),\nonumber \\
p{}_{0}^{\prime} & = & l_{1}+\frac{p_{0}^{2}}{q_{1}-p_{1}},\label{eq:primeparaHN3}\\
p{}_{i}^{\prime} & = & p_{i+1},\qquad(i\geq1),\nonumber \\
l{}_{1}^{\prime} & = & l_{2}+\frac{p_{1}p_{0}^{2}}{q_{1}^{2}-p_{1}^{2}},\nonumber \\
l{}_{i}^{\prime} & = & l_{i+1,}\qquad(i\geq2),\nonumber 
\end{eqnarray}
The difference between the primed and unprimed quantities represents
the step from the $\mu$-th to the $\mu+1$-st level in the RG recursion,
with $0\leq\mu<k$. The recursion for the overall scale-factor $I_{k}$
requires more care, as it depends explicitly on $k$ and we have to
take into account the level $\mu$ at which the factor in front of
the integral in Eq.~(\ref{eq:HN3int_reno}) arises, thereby shifting
$k\to k-\mu$. Thus, 
\begin{equation}
I_{k}^{\prime}=I_{k}\left(q_{1}^{2}-p_{1}^{2}\right)^{-2^{k-\mu-3}}.\label{eq:IprimeparaHN3}
\end{equation}

\subsection{Renormalization Group calculation for the determinant of HNNP and
HN6\label{sub:RG_HNNP}}

Now, we evaluate the most general $2^{k}\times2^{k}$-determinant
for the matrix $L_{k}^{(NP,6)}$ representing either the HNNP or the
HN6 network. Again, we simply provide a description of $L_{k}^{(NP,6)}$
for the case $k=4$: 
\begin{widetext}
\begin{eqnarray}
L_{k}^{(NP,6)} & = & \left[\begin{array}{cccccccccccccccc}
q_{4} & -p_{0} & -l_{1} & -p_{1} & -l_{2} & 0 & -p_{2} & 0 & -2l_{3} & 0 & -p_{2} & 0 & -l_{2} & -p_{1} & -l_{1} & -p_{0}\\
-p_{0} & q_{1} & -p_{0} & 0 & -p_{1} & 0 & 0 & 0 & 0 & 0 & 0 & 0 & 0 & 0 & 0 & 0\\
-l_{1} & -p_{0} & q_{2} & -p_{0} & -l_{1} & 0 & 0 & 0 & -p_{2} & 0 & 0 & 0 & 0 & 0 & 0 & 0\\
-p_{1} & 0 & -p_{0} & q_{1} & -p_{0} & 0 & 0 & 0 & 0 & 0 & 0 & 0 & 0 & 0 & 0 & 0\\
-l_{2} & -p_{1} & -l_{1} & -p_{0} & q_{3} & -p_{0} & -l_{1} & -p_{1} & -l_{2} & 0 & 0 & 0 & 0 & 0 & 0 & 0\\
0 & 0 & 0 & 0 & -p_{0} & q_{1} & -p_{0} & 0 & -p_{1} & 0 & 0 & 0 & 0 & 0 & 0 & 0\\
-p_{2} & 0 & 0 & 0 & -l_{1} & -p_{0} & q_{2} & -p_{0} & -l_{1} & 0 & 0 & 0 & 0 & 0 & 0 & 0\\
0 & 0 & 0 & 0 & -p_{1} & 0 & -p_{0} & q_{1} & -p_{0} & 0 & 0 & 0 & 0 & 0 & 0 & 0\\
-2l_{3} & 0 & -p_{2} & 0 & -l_{2} & -p_{1} & -l_{1} & -p_{0} & q_{4} & -p_{0} & -l_{1} & -p_{1} & -l_{2} & 0 & -p_{2} & 0\\
0 & 0 & 0 & 0 & 0 & 0 & 0 & 0 & -p_{0} & q_{1} & -p_{0} & 0 & -p_{1} & 0 & 0 & 0\\
-p_{2} & 0 & 0 & 0 & 0 & 0 & 0 & 0 & -l_{1} & -p_{0} & q_{2} & -p_{0} & -l_{1} & 0 & 0 & 0\\
0 & 0 & 0 & 0 & 0 & 0 & 0 & 0 & -p_{1} & 0 & -p_{0} & q_{1} & -p_{0} & 0 & 0 & 0\\
-l_{2} & 0 & 0 & 0 & 0 & 0 & 0 & 0 & -l_{2} & -p_{1} & -l_{1} & -p_{0} & q_{3} & -p_{0} & -l_{1} & -p_{1}\\
-p_{1} & 0 & 0 & 0 & 0 & 0 & 0 & 0 & 0 & 0 & 0 & 0 & -p_{0} & q_{1} & -p_{0} & 0\\
-l_{1} & 0 & 0 & 0 & 0 & 0 & 0 & 0 & -p_{2} & 0 & 0 & 0 & -l_{1} & -p_{0} & q_{2} & -p_{0}\\
-p_{0} & 0 & 0 & 0 & 0 & 0 & 0 & 0 & 0 & 0 & 0 & 0 & -p_{1} & 0 & -p_{0} & q_{1}
\end{array}\right]
\end{eqnarray}
\end{widetext}
The meaning of the bare form of the renormalizing parameters ($q_{i,}p_{i},l_{i}$)
is the same as in Sec. \ref{sub:RG_HN3}. In particular, the dark-shaded
links in Fig.~\ref{fig:HN-NP} correspond to $p_{i}$, while the green-shaded ones again refer to $l_{i}$, which may be originally absent but emergent in HNNP. The determinant of $L_{k}^{(NP,6)}$
can be evaluated by using the identity in Eq.~(\ref{eq:gaussint})
to write
\begin{widetext}
\begin{eqnarray}
\left[\det L_{k}^{(NP,6)}\right]^{-\frac{1}{2}} & = & I_{k}\iintop_{-\infty}^{\infty}\frac{dx_{0}dx_{2^{k-1}}}{\pi}\exp\left\{ -q_{k}\left(x_{0}^{2}+x_{2^{k-1}}^{2}\right)+4l_{k-1}x_{0}x_{2^{k-1}}\right\} \dotsintop_{-\infty}^{\infty}\left(\prod_{i=1}^{k-1}\,\prod_{j=1}^{2^{k-i}}\frac{dx_{2^{i-1}(2j-1)}}{\sqrt{\pi}}\right)\label{eq:detL_NP6}\nonumber \\
 &  & \exp\left\{ -\sum_{i=1}^{k-1}q_{i}\sum_{j=1}^{2^{k-i}}x_{2^{i-1}(2j-1)}^{2}+2\sum_{i=1}^{k-2}l_{i}\sum_{j=1}^{2^{k-i}}x_{2^{i-1}(2j-2)}x_{2^{i-1}(2j)}\right.\\
 &  & \quad\left.+2\sum_{i=1}^{k-2}p_{i}\sum_{j=1}^{2^{k-i-1}}\left(x_{2^{i-1}(4j-3)}x_{2^{i-1}(4j)}+x_{2^{i-1}(4j-1)}x_{2^{i-1}(4j-4)}\right)+2p_{0}\sum_{j=1}^{2^{k-1}}x_{2j-1}(x_{2j-2}+x_{2j})\right\} .\nonumber 
\end{eqnarray}
\end{widetext}
To solve $\det L_{k}^{(NP,6)}$ recursively, we again integrate over
all variables $x$ with odd index (i.e., the $i=1$ term in the product):
\begin{widetext}
\begin{eqnarray}
 &  & \prod_{j=1}^{2^{k-2}}\iintop_{-\infty}^{\infty}\frac{dx_{4j-3}dx_{4j-1}}{\pi}\exp\left\{ -q_{1}\left(x_{4j-3}^{2}+x_{4j-1}^{2}\right)+2p_{1}\left(x_{4j-3}x_{4j}+x_{4j-1}x_{4j-4}\right)\right.\nonumber \\
 &  & \qquad\qquad\qquad\qquad\qquad\left.+2p_{0}\left[x_{4j-3}\left(x_{4j-4}+x_{4j-2}\right)+x_{4j-1}\left(x_{4j-2}+x_{4j}\right)\right]\right\} ,\\
 & = & \prod_{j=1}^{2^{k-2}}\frac{1}{q_{1}}\exp\left\{ \frac{\left[p_{1}x_{4j-4}+p_{0}\left(x_{4j}+x_{4j-2}\right)\right]^{2}+\left[p_{1}x_{4j}+p_{0}\left(x_{4j-2}+x_{4j-4}\right)\right]^{2}}{q_{1}}\right\} \nonumber \\
 & = & q_{1}^{-2^{k-2}}\exp\left\{ 2\frac{p_{0}^{2}+p_{1}^{2}}{q_{1}}(x_{0}^{2}+x_{2^{k-1}}^{2})+2\frac{p_{0}^{2}+p_{1}^{2}}{q_{1}}\sum_{i=2}^{k-2}\sum_{j=1}^{2^{k-i-1}}x_{2^{i-1}(2j-1)2}+2\frac{p_{0}^{2}}{q_{1}}\sum_{j=1}^{2^{k-2}}x_{(2j-1)2}^{2}\right.\nonumber \\
 &  & \qquad\qquad\left.2\frac{p_{0}^{2}+p_{0}p_{1}}{q_{1}}\sum_{j=1}^{2^{k-1}}x_{2j-2}x_{2j}+4\frac{p_{0}p_{1}}{q_{1}}\sum_{j=1}^{2^{k-2}}x_{(2j-2)2}x_{(2j)2}\right\} \nonumber 
\end{eqnarray}
\end{widetext}
Substituting back into Eq.~(\ref{eq:detL_NP6}) obtains:
\begin{widetext}
\begin{eqnarray}
&&\left[\det L_{k}^{(NP,6)}\right]^{-\frac{1}{2}} =  I_{k}q_{1}^{-2^{k-2}}\iintop_{-\infty}^{\infty}\frac{dx_{0}dx_{2^{k-1}}}{\pi}\exp\left\{ -\left[q_{k}-2\frac{p_{0}^{2}+p_{1}^{2}}{q_{1}}\right]\left(x_{0}+x_{2^{k-1}}\right)+4l_{k-1}x_{0}x_{2^{k-1}}\right\}\nonumber \\
 &  & \dotsintop_{-\infty}^{\infty}\left(\prod_{i=1}^{k-2}\prod_{j=1}^{2^{k-1-i}}\frac{dx_{2^{i}(2j-1)}}{\sqrt{\pi}}\right)\exp\left\{ -\sum_{i=2}^{k-2}\left(q_{i+1}-2\frac{p_{0}^{2}+p_{1}^{2}}{q_{1}}\right)\sum_{j=1}^{2^{k-1-i}}x_{2^{i}(2j-1)}^{2}-\left(q_{2}-2\frac{p_{0}^{2}}{q_{1}}\right)\sum_{j=1}^{2^{k-2}}x_{2(2j-1)}^{2}\right. \\
 &  & \qquad+2\sum_{i=1}^{k-3}p_{i+1}\sum_{j=1}^{2^{k-i-2}}\left(x_{2^{i}(4j-3)}x_{2^{i}(4j)}+x_{2^{i}(4j-1)}x_{2^{i}(4j-4)}\right)+2\sum_{i=2}^{k-3}l_{i+1}\sum_{j=1}^{k-i-1}x_{2^{i}(2j-2)}x_{2^{i}(2j)}\nonumber \\
 &  & \qquad\left.+2\left(l_{2}+2\frac{p_{0}p_{1}}{q_{1}}\right)\sum_{j=1}^{2^{k-2}}x_{2(2j-2)}x_{2(2j)}+2\left(l_{1}+\frac{p_{0}^{2}+p_{0}p_{1}}{q_{1}}\right)\sum_{j=1}^{2^{k-2}}x_{2(2j-1)}\left(x_{2(2j-2)}+x_{2(2j)}\right)\right\} \nonumber 
\end{eqnarray}
\end{widetext}
Substituting $x_{i}^{\prime}=x_{2i}$, this expression is \emph{identical}
in form with Eq.~(\ref{eq:detL_NP6}), and we can identify 
\begin{eqnarray}
q_{1}^{\prime} & = & q_{2}-2\frac{p_{0}^{2}}{q_{1}},\nonumber \\
q_{i}^{\prime} & = & q_{i+1}-2\frac{p_{0}^{2}+p_{1}^{2}}{q_{1}}\qquad(i\geq2),\nonumber \\
p_{0}^{\prime} & = & l_{1}+\frac{p_{0}^{2}+p_{0}p_{1}}{q_{1}},\label{eq:primeparaHNNP}\\
p_{i}^{\prime} & = & p_{i+1}\qquad\qquad\qquad(i\geq1),\nonumber \\
l_{1}^{\prime} & = & l_{2}+2\frac{p_{0}p_{1}}{q_{1}},\nonumber \\
l_{i}^{\prime} & = & l_{i+1}\qquad\qquad\qquad(i\geq2).\nonumber 
\end{eqnarray}
The difference between the primed and unprimed quantities represents
the step from the $\mu$-th to the $\mu+1$-st level in the RG recursion,
with $0\leq\mu<k$. As for HN3 above, the recursion for the overall
scale-factor $I_{k}$ requires a shift $k\to k-\mu$: 
\begin{equation}
I_{k}^{\prime}=I_{k}\, q_{1}^{-2^{k-\mu-2}}.\label{eq:IprimeparaHNNP}
\end{equation}

\section{RG-Evaluation of Network Laplacians\label{sec:RG-Evaluation-of-Network}}

Here, we use the RG-recursions in Eqs.~(\ref{eq:primeparaHN3}) and
(\ref{eq:primeparaHNNP}) to determine the asymptotic scaling behavior
of the determinant of the network Laplacians for large system sizes.
We begin with the example of a simple line, HN2, which is contained
in either equation, to re-derive some familiar results to demonstrate
the procedure.

\subsection{Simple Example: One-dimensional Lattices\label{sub:Example:-One-dimensional-Lattice}}

The RG allows us to find the secular equation for HN2, a one-dimensional
loop of $N=2^{k}$ sites, as a reference. In that case, all sites
have constant degree $d_{i}=2$. With the RG approach from Sec. \ref{sec:Obtaining-the-RG-Recursions},
we have to solve the recursions either in Eqs.~(\ref{eq:primeparaHN3}-\ref{eq:IprimeparaHN3})
or in Eqs.~(\ref{eq:primeparaHNNP}-\ref{eq:IprimeparaHNNP}) for
the initial conditions on the bare parameters, 
\begin{eqnarray}
I_{k}^{(0)} & = & 1\nonumber \\
q_{i}^{(0)} & = & 2-\lambda\qquad(i\geqslant1)\nonumber \\
p_{0}^{(0)} & = & 1\label{eq:IC_HN2}\\
p_{i}^{(0)} & = & 0\qquad(i\geqslant1)\nonumber \\
l_{i}^{(0)} & = & 0\qquad(i\geqslant1).\nonumber 
\end{eqnarray}
Note that we allowed here for a prospective eigenvalue $\lambda$
subtracted from each diagonal element, as indicated by the eigenvalue
Eq.~(\ref{eq:secular}). In both sets of RG-recursions, the equations
simplify to
\begin{eqnarray}
q_{\mu+1}  =  q_{\mu}-2\frac{p_{\mu}^{2}}{q_{\mu}},\qquad p_{\mu+1} & = & \frac{p_{\mu}^{2}}{q_{\mu}} \label{eq:RG-HN2}
\end{eqnarray}
where we used that $q_{\mu}\equiv q_{i}^{(\mu)}$ and $p_{i}^{(\mu)}=l_{i}^{(\mu)}\equiv0$
for all $i\geq1$ while $p_{\mu}\equiv p_{0}^{(\mu)}$. The recursion
for $I_{k}^{(\mu)}$ in either of Eqs. (\ref{eq:IprimeparaHN3}) or
(\ref{eq:IprimeparaHNNP}) has the formal solution
\begin{equation}
I_{k}^{(\mu)}=\prod_{i=0}^{\mu-1}q_{i}^{-2^{k-2-i}}.\label{eq:Imuk_HN2}
\end{equation}

After the $k-1$-fold application of the RG-recursions in Eq.~(\ref{eq:RG-HN2})
reduces the original $2^{k}\times2^{k}$ matrix -- such as in Eq.
(\ref{eq:H_k^a}) -- down to a $2\times2$ matrix for a ``loop''
with only two sites that are doubly linked, and we have 
\begin{eqnarray}
\det\left[{\bf {\bf L}_{k}^{(2)}}-\lambda{\bf 1}\right] & = & \left[I_{k}^{(k-1)}\right]^{-2}\det\left[\begin{array}{cc}
q_{k-1} & -2p_{k-1}\\
-2p_{k-1} & q_{k-1}
\end{array}\right].\label{eq:HN2LaplDet_RG}
\end{eqnarray}

\paragraph{Exact Solution for HN2:}
The nonlinear recursions in Eq.~(\ref{eq:RG-HN2}) are easily solved
in closed form by defining $s_{\mu}=q_{\mu}/p_{\mu}$, for which $s_{\mu+1}=s_{\mu}^{2}-2$,
obtained by dividing the \emph{2nd} by the \emph{3rd} line. The solution
is 
\begin{equation}
s_{\mu}=2\cos\left[2^{\mu}\arccos\left(\frac{q_{0}}{2p_{0}}\right)\right]=2T_{2^{\mu}}\left(\frac{q_{0}}{2p_{0}}\right)\label{eq:sn}
\end{equation}
where $T_{n}(x)$ refers to the $n$-th Chebyshev polynomial of the
first kind~\cite{abramowitz:64}. Inserting into Eqs.~(\ref{eq:RG-HN2})
and applying the initial conditions in Eqs.~(\ref{eq:IC_HN2}) generates
the results 
\begin{eqnarray}
p_{\mu}  =  \prod_{i=0}^{\mu-1}\frac{1}{s_{i}},\quad
q_{\mu}  =  s_{\mu}\prod_{i=0}^{\mu-1}\frac{1}{s_{i}},
\quad I_{k}^{(\mu)}  =  \prod_{i=0}^{\mu-1}\frac{1}{s_{i}}\equiv p_{\mu},
\label{eq:HN2solutions}
\end{eqnarray}
where the last equality emerges from Eq. (\ref{eq:Imuk_HN2}) under
reordering factors in the products. Alternatively, one could realize
that the Ansatz
\begin{equation}
p_{\mu}=\frac{s\,\zeta^{2^{\mu}}}{1-\zeta^{2^{\mu+1}}},\qquad q_{\mu}=\frac{s\left(1+\zeta^{2^{\mu+1}}\right)}{1-\zeta^{2^{\mu+1}}},
\end{equation}
also provides an exact solution of Eq. (\ref{eq:RG-HN2}), which match
the initial conditions at $\mu=0$ for $s=\pm i\sqrt{\left(4-\lambda\right)\lambda}$
and $\zeta=1-\frac{\lambda}{2}\mp\frac{i}{2}\sqrt{\left(4-\lambda\right)\lambda}$.
In either case, note that for $\lambda\to0$ the solution reduces
to $I_{\mu}=p_{\mu}=q_{\mu}/2=2^{-\mu}$.

We note that these formal solutions, albeit closed-form, are rather
complicated and even numerically very difficult to use for arbitrary
$\lambda$. However, inserting Eq.~(\ref{eq:HN2solutions}) into
Eq. (\ref{eq:HN2LaplDet_RG}) with $N=2^{k}$ provides 
\begin{eqnarray}
\det\left[{\bf {\bf L}_{k}^{(2)}}-\lambda{\bf 1}\right] & = & 2T_{N}\left(1-\frac{\lambda}{2}\right)-2,\label{eq:HN2LaplDet}
\end{eqnarray}
a well-known exact result. Expanding Eq.~(\ref{eq:HN2LaplDet}) to
first order in $\lambda$ provides for the number of spanning ``trees''
in Eq.~(\ref{eq:nST}): 
\begin{equation}
\#_{{\rm ST}}^{(2)}=-\frac{T_{N}^{\prime}(1)}{N}=N,\label{eq:=00003D000023ST_HN2}
\end{equation}
which is exactly the number of open strings covering all sites that
one can embed on a closed loop of $N$ sites.

\paragraph{Asymptotic Solution for HN2:}

In general, we will not be able to solve the nonlinear recursion equations
of the RG-flow in closed form, of course. It is therefore instructive
to explore asymptotic methods for such cases by way of this exactly
solvable instance. We note two properties of the RG-flow that turn
out to be quite general: (1) Evolving numerically from the initial,
bare parameter values in Eq.~(\ref{eq:IC_HN2}) for $\lambda=0$,
these parameters approach their asymptotic value at $\mu\to\infty$
with a correction that decays as $\alpha^{-\mu}$ for some $\alpha>1$;
and (2), to remove the trivial $\lambda=0$ eigenvalue that each of
the network Laplacians ${\bf {\bf L}_{k}}$ possesses, we further
need to expand the RG-flow recursions to first order in $\lambda$
for $\lambda\to0$ while $\mu\to\infty$. As the stable fixed point
for $\mu\to\infty$ at $\lambda=0$ becomes unstable for $\lambda>0$,
that correction diverges with some power $\beta^{\mu}$ for $\beta>1$.
Therefore, we need to make an Ansatz typical to analyze unstable critical
points in RG \cite{SWlong,Pathria}: 
\begin{eqnarray}
p_{\mu} & \sim & p_{\infty}+\alpha^{-\mu}P_{0}+\lambda\beta^{\mu}P_{1}+\ldots,\nonumber \\
q_{\mu} & \sim & q_{\infty}+\alpha^{-\mu}Q_{0}+\lambda\beta^{\mu}Q_{1}+\ldots,\label{eq:Ansatz}
\end{eqnarray}
for $\mu\to\infty$ and $\lambda\to0$. For HN2, $p_{\infty}=q_{\infty}=0$.
Then, we obtain at $\lambda=0$ and to leading order in $\alpha^{-\mu}$:
\begin{equation}
\frac{Q_{0}}{\alpha}=Q_{0}-2\frac{P_{0}^{2}}{Q_{0}},\qquad\frac{P_{0}}{\alpha}=\frac{P_{0}^{2}}{Q_{0}},
\end{equation}
which have the solution 
\begin{equation}
\alpha=2,\qquad Q_{0}=2P_{0}.
\end{equation}
Extending to the order-$\lambda$ correction yields 
\begin{equation}
\beta Q_{1}=\frac{3}{2}Q_{1}-2P_{1},\qquad\beta P_{1}=P_{1}-\frac{1}{4}Q_{1},
\end{equation}
with solution 
\begin{equation}
\beta=2,\qquad Q_{1}=-4P_{1}.
\end{equation}

The first line in Eq.~(\ref{eq:RG-HN2}) in that form is difficult
to evaluate. An asymptotic evaluation doesn't seem possible, generally,
since most contributions arise from the terms with smallest $\mu$
where the power $2^{k-\mu}$ is largest but $q_{\mu}$ in Eq. (\ref{eq:Ansatz})
has not achieved its asymptotic form yet. It is easy to evaluate numerically
to any accuracy, though, when rewritten as a recursion in $k$, $I_{k+1}^{(k)}=\frac{1}{q_{k-1}}\left[I_{k}^{(k-1)}\right]^{2}$.
Here, one easily finds that $I_{k}^{(k-1)}=2^{-k}=1/N$, since $I_{k}^{(k-1)}=p_{k-1}$
for \emph{all} $k$, of course. Here, no $O(\lambda)$-correction
was needed, since the denominator of Eq.~(\ref{eq:HN2LaplDet_RG})
is already finite at $\lambda=0$. In contrast, the numerator cancels
at $\lambda=0$ and yields instead $q_{k-1}^{2}-4p_{k-1}^{2}\sim\lambda\left(\frac{\beta}{\alpha}\right)^{k-1}$
to next order. We finally get for Eq.~(\ref{eq:HN2LaplDet_RG}):
\begin{equation}
\det\left[{\bf {\bf L}_{k}^{(2)}}-\lambda{\bf 1}\right]\sim\lambda\, N^{2+\log_{2}\frac{\beta}{\alpha}},\label{eq:detHN2}
\end{equation}
where we have ignored overall factors. In this simple case, it happens
to be $\alpha=\beta$. Taking these facts into account, Eq.~(\ref{eq:detHN2})
reproduces the exact result in Eq.~(\ref{eq:=00003D000023ST_HN2}).
Note that to obtain the dominant contribution (if it exists) that
varies exponentially in system size $N=2^{k}$, we merely need $I_{k}^{(\mu)}$
for $\lambda=0$; the remaining integral contributes at most a power-law
pre-factor, aside from the $\lambda=0$ trivial eigenvalue itself.

\subsection{Case HN3\label{sub:Case-HN3}}

In this case, we have to interpret the results in Eqs.~(\ref{eq:primeparaHN3})
for the bare parameters 
\begin{eqnarray}
I_{k}^{(0)} & = & 1,\nonumber \\
q_{i}^{(0)} & = & 3-\lambda\qquad(i\geqslant1),\nonumber \\
p_{i}^{(0)} & = & 1\qquad(i\geqslant0),\label{eq:IC_HN3}\\
l_{i}^{(0)} & = & 0\qquad(i\geqslant1),\nonumber 
\end{eqnarray}
reflecting the fact that all sites in HN3 have a constant degree $d_{i}=3$.
Since all diagonal entries are identical, the hierarchy for the $q_{i}$
collapses and we retain only two nontrivial relations, one for $q_{1}$
and one for all other $q_{i}\equiv q_{2}$ for all $i\geq2$. Here,
all $p_{i}$ are non-zero, encompassing the backbone links ($i=0$)
and all levels of long-range links ($i\geq1$). But it remains $p_{i}\equiv1$
for $i\geq1$ at any step $\mu$ of the RG, in particular, $p_{1}^{(\mu)}\equiv1$
throughout; only the backbone $p_{0}$ renormalizes nontrivially.
Although all \emph{bare} links of type $l_{i}$ are absent in this
network, the details of the calculation in Sec. \ref{sub:RG_HN3}
show that under renormalization terms of type $l_{1}$ emerge while
those for $l_{i}$ for $i\geq2$ remain zero at any step. Thus, Eqs.
(\ref{eq:primeparaHN3}-\ref{eq:IprimeparaHN3}) reduce to RG recursion
equations that are far more elaborate than for HN2 in Eq.~(\ref{eq:RG-HN2})
above: 
\begin{eqnarray}
I_{k}^{(\mu+1)} & = & I_{k}^{(\mu)}\left\{ \left[q_{1}^{(\mu)}\right]^{2}-1\right\} ^{-2^{k-3-\mu}},\nonumber \\
q_{1}^{(\mu+1)} & = & q_{2}^{(\mu)}-2\frac{\left[p_{0}^{(\mu)}\right]^{2}}{q_{1}^{(\mu)}-1},\nonumber \\
q_{2}^{(\mu+1)} & = & q_{2}^{(\mu)}-2\frac{q_{1}^{(\mu)}\left[p_{0}^{(\mu)}\right]^{2}}{\left[q_{1}^{(\mu)}\right]^{2}-1},\label{eq:RG-HN3}\\
p_{0}^{(\mu+1)} & = & l_{1}^{(\mu)}+\frac{\left[p_{0}^{(\mu)}\right]^{2}}{q_{1}^{(\mu)}-1},\nonumber \\
l_{1}^{(\mu+1)} & = & \frac{\left[p_{0}^{(\mu)}\right]^{2}}{\left[q_{1}^{(\mu)}\right]^{2}-1}.\nonumber 
\end{eqnarray}
We can further eliminate $q_{2}$ from Eqs.~(\ref{eq:RG-HN3}) by
noting that these equations possess an \emph{invariant}: 
\begin{eqnarray}
q_{2}^{(\mu)} & = & q_{1}^{(\mu)}+2l_{1}^{(\mu)}\qquad(0\leqslant\mu<k).\label{eq:q_invariant}
\end{eqnarray}
Then, abbreviating $q_{\mu}\equiv q_{1}^{(\mu)}$, $p_{\mu}\equiv p_{0}^{(\mu)}$,
and $l_{\mu}=l_{1}^{(\mu)}$, Eqs.~(\ref{eq:RG-HN3}) reduce to 
\begin{eqnarray}
q_{\mu+1} & = & q_{\mu}+2l_{\mu}-2\frac{p_{\mu}^{2}}{q_{\mu}-1}\qquad(q_{0}=3-\lambda),\nonumber \\
p_{\mu+1} & = & l_{\mu}+\frac{p_{\mu}^{2}}{q_{\mu}-1}\qquad(p_{0}=1),\label{eq:RG-HN3_redux}\\
l_{\mu+1} & = & \frac{p_{\mu}^{2}}{q_{\mu}^{2}-1}\qquad(l_{0}=0).\nonumber 
\end{eqnarray}
We can again solve for $I_{k}^{(\mu)}$ in Eq.~(\ref{eq:RG-HN3})
independently, 
\begin{eqnarray}
I_{k}^{(\mu)} & = & \prod_{i=0}^{\mu-1}\left[q_{i}^{2}-1\right]^{-2^{k-3-i}}.\label{eq:Ikproduct}
\end{eqnarray}

Evolving the RG for $k-2$ steps results in a reduced network that
consists of a loop of 4 sites, formerly at $0$, $2^{k-2}$, $2^{k-1}$,
and $3\,2^{k-2}$. The sites at $0$ and at $2^{k-1}$ are now connected
doubly by links $l_{k-2}$ whereas the other two are connected by
a \emph{previously unrenormalized} link of bare unit weight. Each
site is of course connected to its nearest neighbor in the backbone
loop with renormalized links $p_{k-2}$. Note, that the sites at $0$
and at $2^{k-1}$ have a on-site factor of $q_{2}^{(k-2)}=q_{1}^{(k-2)}+2l_{1}^{(k-2)}=q_{k-2}+2l_{k-2}$
whereas the other two sites have a factor of $q_{1}^{(k-2)}=q_{k-2}$.
Hence, the secular determinant for HN3 reads: 
\begin{widetext}
\begin{eqnarray}
\det\left[{\bf {\bf L}_{k}^{(3)}}-\lambda{\bf 1}\right] & = & \frac{1}{\left[I_{k}^{(k-2)}\right]^{2}}\,\det\left[\begin{array}{cccc}
q_{k-2}+2l_{k-2} & -p_{k-2} & -2l_{k-2}-1 & -p_{k-2}\\
-p_{k-2} & q_{k-2} & -p_{k-2} & -1\\
-2l_{k-2}-1 & -p_{k-2} & q_{k-2}+2l_{k-2} & -p_{k-2}\\
-p_{k-2} & -1 & -p_{k-2} & q_{k-2}
\end{array}\right]\label{eq:secularHN3}\\
 & = & \left[I_{k}^{(k-2)}\right]^{-2}\left(q_{k-2}+1\right)\left(q_{k-2}+4l_{k-2}+1\right)\left[(q_{k-2}-1)^{2}-4p_{k-2}^{2}\right]\nonumber 
\end{eqnarray}
\end{widetext}

The evaluation of Eq.~(\ref{eq:secularHN3}) follows closely the
analysis in Sec. \ref{sub:Example:-One-dimensional-Lattice}. Our
numerical investigations indicate that the recursion equations in
(\ref{eq:RG-HN3_redux}), starting from the initial conditions with
$\lambda=0$, evolve to a fixed point at which $q_{\mu}\to q_{\infty}=1$
while $p_{\mu}\sim l_{\mu}\to p_{\infty}=l_{\infty}=0$. Thus, an
Ansatz for fixed points \cite{Pathria} similar to Eq.~(\ref{eq:Ansatz}),
\begin{align}
q_{\mu} & \sim1+\alpha^{-\mu}Q_{0}+\lambda\beta^{\mu}Q_{1},\nonumber \\
p_{\mu} & \sim\alpha^{-\mu}P_{0}+\lambda\beta^{\mu}P_{1},\label{eq:Ansatz-HN3}\\
l_{\mu} & \sim\alpha^{-\mu}L_{0}+\lambda\beta^{\mu}L_{1},\nonumber 
\end{align}
for $\mu\to\infty$ and $\lambda\to0$ and requiring $\alpha,\beta>1$,
when inserted in Eq.~(\ref{eq:RG-HN3_redux}), yields the unique
solutions 
\begin{align}
\alpha & =\frac{2}{\phi},\quad P_{0}=\frac{Q_{0}}{2},\quad L_{0}=\frac{Q_{0}}{4\phi},\label{eq:Ansatz1results}\\
\beta & =2,\quad P_{1}=-\frac{5Q_{1}}{12},\quad L_{1}=-\frac{Q_{1}}{6}.\nonumber 
\end{align}
Here, $\phi=\left(\sqrt{5}+1\right)/2=1.618\ldots$ is the ``golden
ratio''~\cite{Livio03}, and $Q_{0,1}$ remain as arbitrary overall
constants, whose knowledge would require a global solution of Eqs.
(\ref{eq:RG-HN3_redux}).

When applied to Eq.~(\ref{eq:secularHN3}), the determinant becomes
$\sim\lambda\phi^{k-2}\sim\lambda N$. The pre-factor becomes $\left[I_{k}^{(k-2)}\right]^{-2}=\prod_{\mu=0}^{k-3}\left[q_{\mu}^{2}-1\right]^{2^{k-2-\mu}}\sim\prod_{\mu=0}^{k-3}\left(2Q_{0}\alpha^{-\mu}\right)^{2^{k-2-\mu}}\sim x^{N}\alpha^{-\frac{N}{2}+2(k-1)}$
, where $x$ is a constant that can be determined to any accuracy
from Eq. (\ref{eq:Ikproduct}). For instance, it provides the recursion
$I_{k+1}^{(k-1)}=\frac{1}{q_{k-2}^{2}-1}\left[I_{k}^{(k-2)}\right]^{2}$
for $k\geq2$ from which we extract $x\sim\left\{ \left[I_{k}^{(k-2)}\right]^{-2}\right\} ^{1/N}=\left[I_{k}^{(k-2)}\right]^{-2^{1-k}}$
for large $k$, e.g., $x\approx2.0189990298$ at $k=40$. Inserted
into Eq.~(\ref{eq:secularHN3}), we obtain

\begin{eqnarray}
\det\left[{\bf {\bf L}_{k}^{(3)}}-\lambda{\bf 1}\right] & \sim & \lambda N^{2-\log_{2}\phi}x^{N}\label{eq:Ikasymp}
\end{eqnarray}
for $\lambda\to0$, where we have ignored any pre-factors again. Note
the similarity to the calculation leading to Eq.~(\ref{eq:HN2LaplDet}).
By Eq.~(\ref{eq:nST}), we then find for the number of spanning trees
on HN3: 
\begin{equation}
\#_{{\rm ST}}^{(3)}\sim N^{1-\log_{2}\phi}2.0189990298^{N}.\label{eq:ST_HN3}
\end{equation}

\paragraph*{HN3 without backbone:\label{sub:HN3-without-backbone:}}

An extreme check for the consistency of the RG recursions in Eqs.
(\ref{eq:primeparaHN3}-\ref{eq:IprimeparaHN3}) is provided by the
degenerate case of Hanoi networks without backbone. For HN3, this
would imply that the bare parameter equations in (\ref{eq:IC_HN3})
are modified to $q_{i}^{(0)}=1-\lambda$ and $p_{0}^{(0)}=0$. In
that case, all recursions in Eqs.~(\ref{eq:primeparaHN3}-\ref{eq:IprimeparaHNNP})
become trivial, with $p_{0}^{(\mu)}=l_{i}^{(\mu)}=0$ and $q_{i}^{(\mu)}=1-\lambda$
for all $i\geq0$ and $0\leq\mu<k$ while $p_{i}^{(\mu)}=1$ only
for $i\geq1$. Now, since $q_{1}^{2}-p_{1}^{2}\sim2\lambda$ for all
$\mu$, we find $I_{k-2}^{-2}\sim\left(-2\lambda\right)^{\frac{N}{2}-1}$
and that the determinant of the last four sites merely has single
line with $\sim-2\lambda$, such that $\det\left[{\bf {\bf L}_{k}}-\lambda{\bf 1}\right]\sim\left(-2\lambda\right)^{\frac{N}{2}}$.
This correctly reflects the fact that HN3 without backbone decomposes
into $\frac{N}{2}$ disconnected individual links, see Fig.~\ref{fig:HN3},
each link by itself has a trivial determinant $=\left|\begin{array}{cc}
1-\lambda & -1\\
-1 & 1-\lambda
\end{array}\right|\sim-2\lambda$. Clearly, each such determinant divide by the number of its sites
($=2)$ and $\lambda$, according to Eq.~(\ref{eq:nST}), merely
says that single links have a unique spanning tree.

\subsection{Case HN5\label{sub:Case-HN5}}

HN5 contains HN3 but possesses many additional links between sites,
see Fig.~\ref{fig:HN3}, and therefore, we expect that the number
of possible spanning trees proliferates faster in HN5. In this case,
we have to interpret the results in Eq.~(\ref{eq:primeparaHN3})
for the ``bare'' parameters as 
\begin{eqnarray}
I_{k}^{(0)} & = & 1,\nonumber \\
q_{i}^{(0)} & = & 2i+1-\lambda\qquad(i\geqslant1),\label{eq:IC_HN5}\\
p_{i}^{(0)} & = & 1\qquad(i\geqslant0),\nonumber \\
l_{i}^{(0)} & = & 1\qquad(i\geqslant1),\nonumber 
\end{eqnarray}
reflecting the fact that all sites in HN5 have a hierarchy-dependent,
increasing degree of $d_{i}=2i+1$ with average 5. Now, diagonal entries
are no longer identical and we have to modify the equations for the
$q_{i}$ when compared to HN3. Yet, the difference between $q_{i}$
and $q_{i+1}$ is constant throughout, $q_{i+1}-q_{i}=2$, and taking
that modification into account, only the renormalization of $q_{1}$
and $q_{2}$ evolves nontrivially, as before for HN3. Again, all $p_{i}$
are non-zero, encompassing the backbone links ($i=0$) and all levels
of long-range links ($i\geq1$). But it remains $p_{i}\equiv1$ for
$i\geq1$ at any step $\mu$ of the RG-flow, in particular, $p_{1}^{(\mu)}\equiv1$
throughout; only the backbone $p_{0}$ renormalizes nontrivially.
Special to HN5, all links of type $l_{i}$ are already present initially
in this network. Though, only $l_{1}$ renormalizes, as in HN3, whereas
it is $l_{i}\equiv1$ for all $i\geq2$. Thus, we obtain 
more elaborate RG recursion equations which merely differ in the last
relation from Eq.~(\ref{eq:RG-HN3}): 
\begin{eqnarray}
l_{1}^{(\mu+1)} & = & 1+\frac{\left[p_{0}^{(\mu)}\right]^{2}}{\left[q_{1}^{(\mu)}\right]^{2}-1}.\label{eq:RG-HN5}
\end{eqnarray}
The formal solution for $I_{k}^{(\mu)}$ is unchanged from HN3, given
in Eq. (\ref{eq:Ikproduct}). Furthermore, we note that, despite the
changes to the recursions for $q_{2}$ and $l_{1}$, the invariant
in Eq. (\ref{eq:q_invariant}) \emph{remains} valid, allowing the
elimination of the recursion for $q_{2}^{(\mu)}$ for $\mu<k-2$.
Then, abbreviating again $q_{\mu}=q_{1}^{(\mu)}$, $p_{\mu}=p_{0}^{(\mu)}$,
and $l_{\mu}=l_{1}^{(\mu)}$ reduces the RG-recursions to 
\begin{eqnarray}
q_{\mu+1} & = & q_{\mu}+2l_{\mu}-2\frac{p_{\mu}^{2}}{q_{\mu}-1}\qquad(q_{0}=3-\lambda),\nonumber \\
p_{\mu+1} & = & l_{\mu}+\frac{p_{\mu}^{2}}{q_{\mu}-1}\qquad(p_{0}=1),\label{eq:RG-HN5_redux}\\
l_{\mu+1} & = & 1+\frac{p_{\mu}^{2}}{q_{\mu}^{2}-1}\qquad(l_{0}=1).\nonumber 
\end{eqnarray}

As in HN3, evolving the RG for $k-2$ steps also results in a reduced
network that consists of a loop of 4 sites, formerly at $0$, $2^{k-2}$,
$2^{k-1}$, and $3\,2^{k-2}$. The sites at $0$ and at $2^{k-1}$
are now connected doubly by links $l_{k-2}$. Each site is of course
connected to its nearest neighbor in the backbone loop with renormalized
links $p_{k-2}$. Note, since the invariant in Eq. (\ref{eq:q_invariant})
is invalid for $\mu=k-2$ due to degree $d_{k}\neq2k+1$ for sites
$0$ and $2^{k-1}$, special consideration is required for the on-site
factor of $q_{2}^{(k-2)}$, abbreviated as $r_{k-2}$, whereas the
other two sites have a factor of $q_{1}^{(k-2)}=q_{k-2}$. Hence,
the secular determinant for HN5 reads:
\begin{widetext}
\begin{eqnarray}
\det\left[{\bf {\bf L}_{k}^{(5)}}-\lambda{\bf 1}\right] & = & \frac{1}{\left[I_{k}^{(k-2)}\right]^{2}}\,\det\left[\begin{array}{cccc}
r_{k-2} & -p_{k-2} & -2l_{k-2} & -p_{k-2}\\
-p_{k-2} & q_{k-2} & -p_{k-2} & -1\\
-2l_{k-2} & -p_{k-2} & r_{k-2} & -p_{k-2}\\
-p_{k-2} & -1 & -p_{k-2} & q_{k-2}
\end{array}\right]\label{eq:secularHN5}\\
 & = & \left[I_{k}^{(k-2)}\right]^{-2}\,\left(q_{k-2}+1\right)\left(2l_{k-2}+r_{k-2}\right)\left[\left(1-q_{k-2}\right)\left(2l_{k-2}-r_{k-2}\right)-4p_{k-2}^{2}\right]\nonumber 
\end{eqnarray}
\end{widetext}

The seemingly innocuous difference in the last relation of Eqs. (\ref{eq:RG-HN5_redux})
compared to Eq.~(\ref{eq:RG-HN3_redux}) for HN3 has dramatic consequences.
Instead of the singular scaling Ansatz in Eq.~(\ref{eq:Ansatz-HN3})
we used for HN3 (or HN2), Eq.~(\ref{eq:RG-HN5_redux}) has an ordinary
fixed point at $\mu\to\infty$ with $q_{\infty}=(5+\sqrt{41})/2$,
$p_{\infty}=2l_{\infty}=(3+\sqrt{41})/4$ as algebraic solutions at
the fixed point. A simple perturbation for small $\lambda$ on Eq.~(\ref{eq:RG-HN5_redux}) then yields \cite{Pathria} 
\begin{eqnarray}
q_{\mu} & \sim & \frac{5+\sqrt{41}}{2}+\lambda\,2^{\mu}Q_{1},\nonumber \\
p_{\mu} & \sim & \frac{3+\sqrt{41}}{4}-\lambda\,2^{\mu}Q_{1}\frac{57-7\sqrt{41}}{40},\label{eq:Ansatz-HN5}\\
l_{\mu} & \sim & \frac{3+\sqrt{41}}{8}-\lambda\,2^{\mu}Q_{1}\frac{47-7\sqrt{41}}{40}.\nonumber 
\end{eqnarray}
By the same method as described in Sec. \ref{sub:Case-HN3}, we obtain
$x\approx2.7548806715$ at $k=40$ for $\left[I_{k}^{(k-2)}\right]^{-2}\sim x^{N}$.
Inserted into Eq.~(\ref{eq:secularHN5}), we obtain (up to a factor):
\begin{eqnarray}
\det\left[{\bf {\bf L}_{k}^{(5)}}-\lambda{\bf 1}\right] & \sim & \lambda N\,2.7548806715^{N},\label{eq:det_HN5}
\end{eqnarray}
or, for the number of spanning trees, 
\begin{equation}
\#_{{\rm ST}}^{(5)}\sim2.7548806715^{N}.\label{eq:=00003D000023ST_HN5}
\end{equation}

\paragraph*{HN5 without backbone:\label{sub:HN5-without-backbone:}}
As Fig.~\ref{fig:HN3} suggests, HN5 without its backbone implies
$p_{0}^{(0)}=0$ but $p_{i}^{(0)}=l_{i}^{(0)}=1$ and $q_{i}^{(0)}=2i-1-\lambda$
for all $i\geq1$. Then, the \emph{first} iteration of the RG recursions
in Eqs.~(\ref{eq:primeparaHN3}-\ref{eq:IprimeparaHN3}) evolves
trivially, producing $p_{0}^{(1)}=1$ and $q_{i}^{(1)}=q_{i+1}^{(0)}=2i+1-\lambda$
as well as $p_{i}^{(1)}=l_{i}^{(1)}=1$ for all $i\geq1$. If we also
enforce $I_{k}^{(1)}=1$, these are just the initial conditions of
the RG-recursions in Eq.~(\ref{eq:IC_HN5}) again, except starting
at $\mu=1$, and we reproduce the same result as in Eqs.~(\ref{eq:det_HN5}-\ref{eq:=00003D000023ST_HN5})
but for a network of size $N/2$. One complication arises from the
$N/4$ disconnected $p_{1}$-links. These are accounted for via the
first iteration of $I_{k}^{(\mu)}$ in Eq.~(\ref{eq:IprimeparaHN3}):
with $q_{1}^{(0)}=1-\lambda$ and $p_{1}^{(0)}=1$, it produces a
factor of $\left[I_{k}^{(1)}\right]^{-2}=\left[q_{1}^{2}-p_{1}^{2}\right]^{2^{k-2}}\sim\left(-2\lambda\right)^{\frac{N}{4}}$,
i.e., exactly one factor of $-2\lambda$ for each disconnected line.
In the calculation of the spanning trees of the remaining network,
these need to be ignored, hence, requiring $I_{k}^{(1)}=1$.

\subsection{Case HNNP}

In this case, we have to interpret the results in Eqs.~(\ref{eq:primeparaHNNP}-\ref{eq:IprimeparaHNNP})
for the bare parameters
\begin{eqnarray}
I_{k}^{(0)} & = & 1,\nonumber \\
q_{1}^{(0)} & = & 3-\lambda,\nonumber \\
q_{i}^{(0)} & = & 2i-1-\lambda\hspace{1em}(i\geqslant2),\label{eq:HNNP_IC}\\
p_{i}^{(0)} & = & 1\qquad\quad(i\geqslant1),\nonumber \\
l_{i}^{(0)} & = & 0\qquad\quad(i\geqslant1),\nonumber 
\end{eqnarray}
reflecting the fact that all sites in HNNP have a hierarchy-dependent,
increasing degree of $d_{i}=2i-1$, $2\leqslant i<k$, as for HN5
above, but instead with average degree 4. Here, too, the difference
between $q_{i}$ and $q_{i+1}$ for $i\geq2$ is constant throughout,
$q_{i+1}-q_{i}=2$, such that only the renormalization of $q_{1}$
and $q_{2}$ evolve nontrivially. Again, it remains $p_{i}\equiv1$
for $i\geqslant1$ at any step $\mu$ of the RG, in particular, $p_{1}^{(\mu)}\equiv1$
throughout and only the backbone $p_{0}$ renormalizes nontrivially.
Although all links of type $l_{i}$ are initially absent in this network,
under renormalization terms of type $l_{1}$ emerge while those for
$l_{i}$ for $i\geqslant2$ remain zero at any step. These considerations
reduce Eqs.~(\ref{eq:primeparaHNNP}-\ref{eq:IprimeparaHNNP}) to
\begin{eqnarray}
I_{k}^{(\mu+1)} & = & I_{k}^{(\mu)}[q_{1}^{(\mu)}]^{-2^{(k-\mu-2)}},\nonumber \\
q_{1}^{(\mu+1)} & = & q_{2}^{(\mu)}-2\frac{[p_{0}^{(\mu)}]^{2}}{q_{1}^{(\mu)}},\nonumber \\
q_{2}^{(\mu+1)} & = & q_{2}^{(\mu)}+2-2\frac{[p_{0}^{(\mu)}]^{2}+1}{q_{1}^{(\mu)}},\label{eq:HNNP-RG}\\
p_{0}^{(\mu+1)} & = & l_{1}^{(\mu)}+\frac{[p_{0}^{(\mu)}]^{2}+p_{0}^{(\mu)}}{q_{1}^{(\mu)}},\nonumber \\
l_{1}^{(\mu+1)} & = & 2\frac{p_{0}^{(\mu)}}{q_{1}^{(\mu)}}.\nonumber 
\end{eqnarray}

Then, abbreviating $q_{\mu}\equiv q_{1}^{(\mu)}$, $r{}_{\mu}\equiv q_{2}^{(\mu)}$,
$p_{\mu}\equiv p_{0}^{(\mu)}$, and $l_{\mu}=l_{1}^{(\mu)}$, Eqs.(\ref{eq:HNNP-RG})
further simplify to 
\begin{eqnarray}
q_{\mu+1} & = & r{}_{\mu}-2\frac{p_{\mu}^{2}}{q_{\mu}},\nonumber \\
r{}_{\mu+1} & = & r_{\mu}+2-2\frac{p_{\mu}^{2}+1}{q_{\mu}},\label{eq:RGflow_HNNP}\\
p_{\mu+1} & = & l_{\mu}+\frac{p_{\mu}^{2}+p_{\mu}}{q_{\mu}},\nonumber \\
l_{\mu+1} & = & 2\frac{p_{\mu}}{q_{\mu}}.\nonumber 
\end{eqnarray}

Evolving the RG for $k-2$ steps results in a reduced network that
consists of 4 sites, formerly in $0,$ $2^{k-2},$ $2^{k-1},$ $3\thinspace2^{k-2}$.
Hence the determinant for HNNP reads
\begin{widetext}
\begin{eqnarray}
\det\left[\mathbf{L^{(NP)}}-\lambda\mathbf{1}\right] & = & \frac{1}{\left[I_{k}^{(k-2)}\right]^{2}}\,\det\left[\begin{array}{cccc}
r_{k-2} & -p_{k-2} & -2l_{k-2} & -p_{k-2}\\
-p_{k-2} & q_{k-2} & -p_{k-2} & 0\\
-2l_{k-2} & -p_{k-2} & r_{k-2} & -p_{k-2}\\
-p_{k-2} & 0 & -p_{k-2} & q_{k-2}
\end{array}\right]\label{eq:secular-HNNP}\\
 & = & \left[I_{k}^{(k-2)}\right]^{-2}\left\{ -q_{k-2}\left(2l_{k-2}+r_{k-2}\right)\left[4p_{k-2}^{2}+q_{k-2}\left(2l_{k-2}-r_{k-2}\right)\right]\right\} .\nonumber 
\end{eqnarray}
\end{widetext}

For HNNP, Eq.(\ref{eq:RGflow_HNNP}) has a fixed point at $\mu\rightarrow\infty$,
with $q_{\infty}=5,r{}_{\infty}=33/5,p_{\infty}=2,l_{\infty}=4/5$.
A simple perturbation for small $\lambda$ on Eq.~(\ref{eq:RGflow_HNNP})
then yields \cite{Pathria}

\begin{align}
q_{\mu} & \sim5+\lambda2^{\mu}Q_{1},\nonumber \\
r_{\mu} & \sim\frac{33}{5}+\lambda2^{\mu}Q_{1}\frac{26}{25}\label{eq:HNNP-RG-FP}\\
p_{\mu} & \sim2-\lambda2^{\mu}Q_{1}\frac{2}{5},\nonumber \\
l_{\mu} & \sim\frac{4}{5}-\lambda2^{\mu}Q_{1}\frac{4}{25}.\nonumber 
\end{align}
Here, $I_{k}^{(\mu)}$ in Eq. (\ref{eq:HNNP-RG}) satisfies the recursion,
$I_{k+1}^{(k-1)}=\frac{1}{q_{k-2}^{2}}\left[I_{k}^{(k-2)}\right]^{2}$,
which yields $x\approx2.949008159$ at $k=40$ for $\left[I_{k}^{(k-2)}\right]^{-2}\sim x^{N}$.
Inserted into Eq.~(\ref{eq:secular-HNNP}), we obtain (up to a factor):
\begin{eqnarray}
\det\left[{\bf {\bf L}_{k}^{(NP)}}-\lambda{\bf 1}\right] & \sim & \lambda N\thinspace2.949008159^{N},\label{eq:det_HNNP}
\end{eqnarray}
or, for the number of spanning trees in HNNP: 
\begin{equation}
\#_{{\rm ST}}^{(NP)}\sim2.949008159^{N}.\label{eq:=00003D000023ST_HNNP}
\end{equation}

\paragraph*{HNNP without backbone:\label{sub:HNNP-without-backbone:} }

This case is interesting because HNNP without its backbone looses
all its loops and decomposes into a fixed number (here, four) of separated
but extensive trees~\cite{Hasegawa13c}. As Eq.~(\ref{eq:primeparaHNNP})
suggests, starting with $p_{0}^{(0)}=l_{i}^{(0)}=0$ and $p_{i}^{(0)}=1$
for all $i\geq1$ implies that $p_{0}^{(\mu)}=l_{i}^{(\mu)}=0$ and
$p_{i}^{(\mu)}=1$ for all $\mu$ and $i\geq1$. Then, using the fact
that $q_{i+1}^{(\mu)}-q_{i}^{(\mu)}=2$ throughout for $i\geq2$,
Eq.~(\ref{eq:primeparaHNNP}) collapses to 
\begin{align}
q_{1}^{(\mu+1)} & =q_{2}^{(\mu)},\label{eq:HNNP_tree}\\
q_{2}^{(\mu+1)} & =q_{2}^{(\mu)}+2-\frac{2}{q_{1}^{(\mu)}}.\nonumber 
\end{align}
Note that only on-site parameters renormalize, as can be expected
for a tree. While these recursions are non-trivial, the initial conditions
$q_{1,2}^{(0)}=1-\lambda$, simply lead to $q_{1,2}^{(\mu)}=1$ for
all $\mu$ when $\lambda=0$. Hence, so is $I_{k}^{(\mu)}=1$ for
all $\mu$.

Including $O(\lambda)$-corrections, Eq.~(\ref{eq:HNNP_tree}) yields
$q_{1,2}^{(\mu)}\sim1+Q_{1,2}2^{\mu}\lambda$ for large $\mu$. In
the final step of the RG, these trees reduce to four \emph{isolated}
sites, at $n=0$, $2^{k-2}$, $2^{k-1}$, and $3\,2^{k-2}$, so that
its Laplacian determinant merely has non-zero diagonal elements of
the form $q_{1,2}^{(k-2)}-1\sim\lambda\frac{N}{4}$, where the $-1$
arises from the fact that these four sites each initially had one
less link than expected from their level in the hierarchy. Thus, $\det\left[{\bf {\bf L}_{k}}-\lambda{\bf 1}\right]\sim\left(\lambda\frac{N}{4}\right)^{4}$,
reflecting the expectation that each of the four disconnected trees
merely contributes a unit factor to the count of spanning trees.

\subsection{Case HN6}

In this case, we have to interpret the results in Eqs.~(\ref{eq:primeparaHNNP})
for the bare parameters

\begin{eqnarray}
I_{k}^{(0)} & = & 1\nonumber \\
q_{1}^{(0)} & = & 3-\lambda\nonumber \\
q_{i}^{(0)} & = & 4i-3-\lambda\hspace{1em}(i\geqslant2)\label{eq:HN6-IC}\\
p_{i}^{(0)} & = & 1\qquad\quad(i\geqslant1)\nonumber \\
l_{i}^{(0)} & = & 1\qquad\quad(i\geqslant1)\nonumber 
\end{eqnarray}
reflecting the fact that all sites in HN6 have a hierarchy-dependent,
increasing degree of $d_{i}=4i-3(2\leqslant i<k)$ with average 6.
The difference between $q_{i}$ and $q_{i+1}$ is constant throughout,
here $q_{i+1}-q_{i}=4$. Only the renormalization of $q_{1}$ and
$q_{2}$ evolve nontrivially, as before. Again, all $p_{i}$ are non-zero,
encompassing the backbone links ($i=0$) and all levels of long-range
links ($i\geqslant1$). But it remains $p_{i}\equiv1$ for $i\geqslant1$
at any step $\mu$ of the RG, in particular, $p_{1}^{(\mu)}\equiv1$
throughout; only the backbone $p_{0}$ renormalizes nontrivially.
As for HN5, in HN6 bare links of type $l_{i}$ are present in this
network. Though, only $l_{1}$ renormalizes, as in HN5, while $l_{i}\equiv1$
remains unrenormalized for all $i\geq2$. Applying these considerations
to Eqs.~(\ref{eq:primeparaHNNP}-\ref{eq:IprimeparaHNNP}) results
in: 
\begin{eqnarray}
I_{k}^{(\mu+1)} & = & I_{k}^{(\mu)}\left[q_{1}^{(\mu)}\right]{}^{-2^{(k-\mu-2)}},\nonumber \\
q_{1}^{(\mu+1)} & = & q_{2}^{(\mu)}-2\frac{\left[p_{0}^{(\mu)}\right]{}^{2}}{q_{1}^{(\mu)}},\nonumber \\
q_{2}^{(\mu+1)} & = & q_{2}^{(\mu)}+4-2\frac{\left[p_{0}^{(\mu)}\right]{}^{2}+1}{q_{1}^{(\mu)}},\label{eq:HN6-RG}\\
p_{0}^{(\mu+1)} & = & l_{1}^{(\mu)}+\frac{\left[p_{0}^{(\mu)}\right]{}^{2}+p_{0}^{(\mu)}}{q_{1}^{(\mu)}},\nonumber \\
l_{1}^{(\mu+1)} & = & 1+2\frac{p_{0}^{(\mu)}}{q_{1}^{(\mu)}}.\nonumber 
\end{eqnarray}
Then, abbreviating $q_{\mu}\equiv q_{1}^{(\mu)}$, $r{}_{\mu}\equiv q_{2}^{(\mu)}$,
$p_{\mu}\equiv p_{0}^{(\mu)}$, and $l_{\mu}=l_{1}^{(\mu)}$, Eqs.
(\ref{eq:HN6-RG}) reduce to 
\begin{eqnarray}
q_{\mu+1} & = & r_{\mu}-2\frac{p_{\mu}^{2}}{q_{\mu}}\nonumber \\
r_{\mu+1} & = & r_{\mu}+4-2\frac{p_{\mu}^{2}+1}{q_{\mu}}\label{eq:HN6-RG-flow}\\
p_{\mu+1} & = & l_{\mu}+\frac{p_{\mu}^{2}+p_{\mu}}{q_{\mu}}\nonumber \\
l_{\mu+1} & = & 1+2\frac{p_{\mu}}{q_{\mu}}.\nonumber 
\end{eqnarray}
For HN6, Eq.~(\ref{eq:HN6-RG-flow}) has a fixed point at $\mu\rightarrow\infty$
with $q_{\infty}=5+2\sqrt{5}$, $r_{\infty}=7+14/\sqrt{5}$, $p_{\infty}=2+\sqrt{5}$,
and $l_{\infty}=1+2/\sqrt{5}$. A simple perturbation for small $\lambda$
on Eq.~(\ref{eq:HN6-RG-flow}) then yields \cite{Pathria}

\begin{align}
q_{\mu} & \sim5+2\sqrt{5}+\lambda2^{\mu}Q_{1},\label{eq:FP_HN6}\\
r_{\mu} & \sim7+\frac{14}{\sqrt{5}}+\lambda2^{\mu}Q_{1}\frac{14-4\sqrt{5}}{5},\nonumber \\
p_{\mu} & \sim2+\sqrt{5}-\lambda2^{\mu}Q_{1}\frac{10-3\sqrt{5}}{10},\nonumber \\
l_{\mu} & \sim1+\frac{2}{\sqrt{5}}-\lambda2^{\mu}Q_{1}\frac{12-5\sqrt{5}}{10}.\nonumber 
\end{align}
Again, $I_{k}^{(\mu)}$ in Eq. (\ref{eq:HN6-RG}) satisfies the recursion,
$I_{k+1}^{(k-1)}=\frac{1}{q_{k-2}^{2}}\left[I_{k}^{(k-2)}\right]^{2}$,
which yields $x\approx4.0977251445$ at $k=40$ for $\left[I_{k}^{(k-2)}\right]^{-2}\sim x^{N}$.
Inserted into Eq.~(\ref{eq:secular-HNNP}), which remains formally
valid for HN6, we obtain (up to a factor): 
\begin{eqnarray}
\det\left[{\bf {\bf L}_{k}^{(6)}}-\lambda{\bf 1}\right] & \sim & \lambda N\thinspace4.0977251445^{N},\label{eq:det_HN6}
\end{eqnarray}
or, for the exponential proliferation of spanning trees, 
\begin{equation}
\#_{{\rm ST}}^{(6)}\sim4.0977251445^{N}.\label{eq:=00003D000023ST_HN6}
\end{equation}

\section{Conclusions\label{sec:Conclusions}}

We have presented a RG procedure to obtain spectral properties of
lattice Laplacians. We have applied this procedure to the exactly
renormalizable Hanoi networks, for which we obtain a rich set of recursion
equations that lend themselves to many potential applications~\cite{Shanshan15}.
Here, we have analyzed these equations to count the asymptotic growth
of spanning trees on these networks, and we have checked their validity
for various extreme limits where they reproduce the expected results.
Generally, the addition of extra links results quite naturally in
an increase of complexity, measured in terms of the entropy-density
defined in Eq.~(\ref{eq:STentropy}), such as when we progress from
HN2 to HN3 and to HN5, or from HN2 to HNNP and to HN6. However, it
is notable that HNNP, merely an average degree-4 network, has a higher
complexity than HN5. Each represents a seemingly small variation in
the basic design pattern of their hierarchical structure. Yet, it
renders HNNP and HN6 non-planar while HN3 and HN5 remain planar, which
may explain the slower growth of the later with respect to the former.
HN3 possesses the weakest growth. It is the only one with a small,
regular degree of sites and with an average distance between sites
that grows faster than logarithmic with system size ($\sim\sqrt{N}$).
Both facts help suppress the combinatorial proliferation of alternative
paths between sites, in comparison with, say, a random regular graph
of degree 3 that is locally tree-like, which has $s=\ln\frac{4}{\sqrt{3}}=0.836988\ldots$~\cite{Lyons05}.
Since the number of spanning trees is a metric of well-connectedness
of networks, dynamical properties such as synchronizability is anticipated
to be progressively better for those Hanoi networks of higher degree,
with an additional advantage for those that are non-planar. 

\begin{table}[b!]
\protect\caption{\label{tab:Entropy-densities-for-spanning}Entropy-densities from
Eq. (\ref{eq:STentropy}) for spanning trees on Hanoi networks.}
\begin{tabular}{|c|r@{\extracolsep{0pt}.}l|}
\hline 
Network & \multicolumn{2}{c|}{$s=\frac{1}{N}\ln\left(\#_{ST}\right)$}\tabularnewline
\hline 
\hline 
HN2 & \multicolumn{2}{c|}{0}\tabularnewline
\hline 
HN3 & 0&7026018588\tabularnewline
\hline 
HN5 & 1&01337412\tabularnewline
\hline 
HNNP & 1&0814688965\tabularnewline
\hline 
HN6 & 1&4104319769\tabularnewline
\hline 
\end{tabular}
\end{table}

\bibliographystyle{apsrev}
\bibliography{/Users/stb/Boettcher,newRefs}

\begin{thebibliography}{57}
\expandafter\ifx\csname natexlab\endcsname\relax\def\natexlab#1{#1}\fi
\expandafter\ifx\csname bibnamefont\endcsname\relax
  \def\bibnamefont#1{#1}\fi
\expandafter\ifx\csname bibfnamefont\endcsname\relax
  \def\bibfnamefont#1{#1}\fi
\expandafter\ifx\csname citenamefont\endcsname\relax
  \def\citenamefont#1{#1}\fi
\expandafter\ifx\csname url\endcsname\relax
  \def\url#1{\texttt{#1}}\fi
\expandafter\ifx\csname urlprefix\endcsname\relax\def\urlprefix{URL }\fi
\providecommand{\bibinfo}[2]{#2}
\providecommand{\eprint}[2][]{\url{#2}}

\bibitem[{\citenamefont{Trusina et~al.}(2004)\citenamefont{Trusina, Maslov,
  Minnhagen, and Sneppen}}]{Trusina04}
\bibinfo{author}{\bibfnamefont{A.}~\bibnamefont{Trusina}},
  \bibinfo{author}{\bibfnamefont{S.}~\bibnamefont{Maslov}},
  \bibinfo{author}{\bibfnamefont{P.}~\bibnamefont{Minnhagen}},
  \bibnamefont{and} \bibinfo{author}{\bibfnamefont{K.}~\bibnamefont{Sneppen}},
  \bibinfo{journal}{Phys. Rev. Lett.} \textbf{\bibinfo{volume}{92}},
  \bibinfo{pages}{178702} (\bibinfo{year}{2004}).

\bibitem[{\citenamefont{Clauset et~al.}(2008)\citenamefont{Clauset, Moore, and
  Newman}}]{Clauset08}
\bibinfo{author}{\bibfnamefont{A.}~\bibnamefont{Clauset}},
  \bibinfo{author}{\bibfnamefont{C.}~\bibnamefont{Moore}}, \bibnamefont{and}
  \bibinfo{author}{\bibfnamefont{M.~E.~J.} \bibnamefont{Newman}},
  \bibinfo{journal}{Nature} \textbf{\bibinfo{volume}{453}}, \bibinfo{pages}{98}
  (\bibinfo{year}{2008}).

\bibitem[{\citenamefont{Fischer et~al.}(2008)\citenamefont{Fischer, Hoffmann,
  and Sibani}}]{Fischer08}
\bibinfo{author}{\bibfnamefont{A.}~\bibnamefont{Fischer}},
  \bibinfo{author}{\bibfnamefont{K.~H.} \bibnamefont{Hoffmann}},
  \bibnamefont{and} \bibinfo{author}{\bibfnamefont{P.}~\bibnamefont{Sibani}},
  \bibinfo{journal}{Phys. Rev. E} \textbf{\bibinfo{volume}{77}},
  \bibinfo{pages}{041120} (\bibinfo{year}{2008}).

\bibitem[{\citenamefont{Boettcher
  et~al.}(2008{\natexlab{a}})\citenamefont{Boettcher, Gon{\c c}alves, and
  Guclu}}]{SWPRL}
\bibinfo{author}{\bibfnamefont{S.}~\bibnamefont{Boettcher}},
  \bibinfo{author}{\bibfnamefont{B.}~\bibnamefont{Gon{\c c}alves}},
  \bibnamefont{and} \bibinfo{author}{\bibfnamefont{H.}~\bibnamefont{Guclu}},
  \bibinfo{journal}{J. Phys. A: Math. Theor.} \textbf{\bibinfo{volume}{41}},
  \bibinfo{pages}{252001} (\bibinfo{year}{2008}{\natexlab{a}}).

\bibitem[{\citenamefont{Hasegawa and Nemoto}(2013)}]{Hasegawa13b}
\bibinfo{author}{\bibfnamefont{T.}~\bibnamefont{Hasegawa}} \bibnamefont{and}
  \bibinfo{author}{\bibfnamefont{K.}~\bibnamefont{Nemoto}},
  \bibinfo{journal}{Phys. Rev. E} \textbf{\bibinfo{volume}{88}},
  \bibinfo{pages}{062807} (\bibinfo{year}{2013}).

\bibitem[{\citenamefont{Simon}(1962)}]{Simon62}
\bibinfo{author}{\bibfnamefont{H.~A.} \bibnamefont{Simon}},
  \bibinfo{journal}{Proc. of the American Philosophical Society}
  \textbf{\bibinfo{volume}{106}}, \bibinfo{pages}{467} (\bibinfo{year}{1962}).

\bibitem[{\citenamefont{Pathria}(1996)}]{Pathria}
\bibinfo{author}{\bibfnamefont{R.~K.} \bibnamefont{Pathria}},
  \emph{\bibinfo{title}{Statistical Mechanics, 2nd Ed.}}
  (\bibinfo{publisher}{Butterworth-Heinemann}, \bibinfo{year}{1996}).

\bibitem[{\citenamefont{{M\'ezard} and Montanari}(2006)}]{Mezard06}
\bibinfo{author}{\bibfnamefont{M.}~\bibnamefont{{M\'ezard}}} \bibnamefont{and}
  \bibinfo{author}{\bibfnamefont{A.}~\bibnamefont{Montanari}},
  \emph{\bibinfo{title}{Constraint Satisfaction Networks in Physics and
  Computation}} (\bibinfo{publisher}{Oxford University Press},
  \bibinfo{address}{Oxford}, \bibinfo{year}{2006}).

\bibitem[{\citenamefont{Dorogovtsev et~al.}(2008)\citenamefont{Dorogovtsev,
  Goltsev, and Mendes}}]{Dorogovtsev08}
\bibinfo{author}{\bibfnamefont{S.~N.} \bibnamefont{Dorogovtsev}},
  \bibinfo{author}{\bibfnamefont{A.~V.} \bibnamefont{Goltsev}},
  \bibnamefont{and} \bibinfo{author}{\bibfnamefont{J.~F.~F.}
  \bibnamefont{Mendes}}, \bibinfo{journal}{Rev. Mod. Phys.}
  \textbf{\bibinfo{volume}{80}}, \bibinfo{pages}{1275} (\bibinfo{year}{2008}).

\bibitem[{\citenamefont{Barthelemy}(2011)}]{barthelemy_spatial_2010}
\bibinfo{author}{\bibfnamefont{M.}~\bibnamefont{Barthelemy}},
  \bibinfo{journal}{Physics Reports} \textbf{\bibinfo{volume}{499}},
  \bibinfo{pages}{1} (\bibinfo{year}{2011}).

\bibitem[{\citenamefont{{Moretti} and {Mu{\~n}oz}}(2013)}]{Moretti13}
\bibinfo{author}{\bibfnamefont{P.}~\bibnamefont{{Moretti}}} \bibnamefont{and}
  \bibinfo{author}{\bibfnamefont{M.~A.} \bibnamefont{{Mu{\~n}oz}}},
  \bibinfo{journal}{Nat. Comm.} \textbf{\bibinfo{volume}{4}},
  \bibinfo{pages}{2521} (\bibinfo{year}{2013}).

\bibitem[{\citenamefont{Meunier et~al.}(2009)\citenamefont{Meunier, Lambiotte,
  Fornito, Ersche, and Bullmore}}]{Meunier09}
\bibinfo{author}{\bibfnamefont{D.}~\bibnamefont{Meunier}},
  \bibinfo{author}{\bibfnamefont{R.}~\bibnamefont{Lambiotte}},
  \bibinfo{author}{\bibfnamefont{A.}~\bibnamefont{Fornito}},
  \bibinfo{author}{\bibfnamefont{K.}~\bibnamefont{Ersche}}, \bibnamefont{and}
  \bibinfo{author}{\bibfnamefont{E.~T.} \bibnamefont{Bullmore}},
  \bibinfo{journal}{Frontiers in Neuroinformatics}
  \textbf{\bibinfo{volume}{3}}, \bibinfo{pages}{37} (\bibinfo{year}{2009}).

\bibitem[{\citenamefont{Hinczewski and Berker}(2006)}]{Hinczewski06}
\bibinfo{author}{\bibfnamefont{M.}~\bibnamefont{Hinczewski}} \bibnamefont{and}
  \bibinfo{author}{\bibfnamefont{A.~N.} \bibnamefont{Berker}},
  \bibinfo{journal}{Phys. Rev. E} \textbf{\bibinfo{volume}{73}},
  \bibinfo{pages}{066126} (\bibinfo{year}{2006}).

\bibitem[{\citenamefont{Boettcher et~al.}(2009)\citenamefont{Boettcher, Cook,
  and Ziff}}]{Boettcher09c}
\bibinfo{author}{\bibfnamefont{S.}~\bibnamefont{Boettcher}},
  \bibinfo{author}{\bibfnamefont{J.~L.} \bibnamefont{Cook}}, \bibnamefont{and}
  \bibinfo{author}{\bibfnamefont{R.~M.} \bibnamefont{Ziff}},
  \bibinfo{journal}{Phys. Rev. E} \textbf{\bibinfo{volume}{80}},
  \bibinfo{pages}{041115} (\bibinfo{year}{2009}).

\bibitem[{\citenamefont{{Boettcher} and {Brunson}}(2011)}]{Boettcher10c}
\bibinfo{author}{\bibfnamefont{S.}~\bibnamefont{{Boettcher}}} \bibnamefont{and}
  \bibinfo{author}{\bibfnamefont{C.~T.} \bibnamefont{{Brunson}}},
  \bibinfo{journal}{Phys. Rev. E} \textbf{\bibinfo{volume}{83}},
  \bibinfo{pages}{021103} (\bibinfo{year}{2011}).

\bibitem[{\citenamefont{Boettcher and Brunson}(2015)}]{BoBr12}
\bibinfo{author}{\bibfnamefont{S.}~\bibnamefont{Boettcher}} \bibnamefont{and}
  \bibinfo{author}{\bibfnamefont{C.~T.} \bibnamefont{Brunson}},
  \bibinfo{journal}{EPL (Europhysics Letters)} \textbf{\bibinfo{volume}{110}},
  \bibinfo{pages}{26005} (\bibinfo{year}{2015}).

\bibitem[{\citenamefont{Nogawa et~al.}(2012)\citenamefont{Nogawa, Hasegawa, and
  Nemoto}}]{PhysRevLett.108.255703}
\bibinfo{author}{\bibfnamefont{T.}~\bibnamefont{Nogawa}},
  \bibinfo{author}{\bibfnamefont{T.}~\bibnamefont{Hasegawa}}, \bibnamefont{and}
  \bibinfo{author}{\bibfnamefont{K.}~\bibnamefont{Nemoto}},
  \bibinfo{journal}{Phys. Rev. Lett.} \textbf{\bibinfo{volume}{108}},
  \bibinfo{pages}{255703} (\bibinfo{year}{2012}).

\bibitem[{\citenamefont{Boettcher et~al.}(2012)\citenamefont{Boettcher, Singh,
  and Ziff}}]{Boettcher11d}
\bibinfo{author}{\bibfnamefont{S.}~\bibnamefont{Boettcher}},
  \bibinfo{author}{\bibfnamefont{V.}~\bibnamefont{Singh}}, \bibnamefont{and}
  \bibinfo{author}{\bibfnamefont{R.~M.} \bibnamefont{Ziff}},
  \bibinfo{journal}{Nature Communications} \textbf{\bibinfo{volume}{3}},
  \bibinfo{pages}{787} (\bibinfo{year}{2012}).

\bibitem[{\citenamefont{Li and Boettcher}(in preparation)}]{Shanshan15}
\bibinfo{author}{\bibfnamefont{S.}~\bibnamefont{Li}} \bibnamefont{and}
  \bibinfo{author}{\bibfnamefont{S.}~\bibnamefont{Boettcher}}
  (\bibinfo{year}{in preparation}).

\bibitem[{\citenamefont{Agliari
  et~al.}(2015{\natexlab{a}})\citenamefont{Agliari, Barra, Galluzzi, Guerra,
  Tantari, and Tavani}}]{agliari15retrieval}
\bibinfo{author}{\bibfnamefont{E.}~\bibnamefont{Agliari}},
  \bibinfo{author}{\bibfnamefont{A.}~\bibnamefont{Barra}},
  \bibinfo{author}{\bibfnamefont{A.}~\bibnamefont{Galluzzi}},
  \bibinfo{author}{\bibfnamefont{F.}~\bibnamefont{Guerra}},
  \bibinfo{author}{\bibfnamefont{D.}~\bibnamefont{Tantari}}, \bibnamefont{and}
  \bibinfo{author}{\bibfnamefont{F.}~\bibnamefont{Tavani}},
  \bibinfo{journal}{Phys. Rev. Lett.} \textbf{\bibinfo{volume}{114}},
  \bibinfo{pages}{028103} (\bibinfo{year}{2015}{\natexlab{a}}).

\bibitem[{\citenamefont{Agliari
  et~al.}(2015{\natexlab{b}})\citenamefont{Agliari, Barra, Galluzzi, Guerra,
  Tantari, and Tavani}}]{agliari15topological}
\bibinfo{author}{\bibfnamefont{E.}~\bibnamefont{Agliari}},
  \bibinfo{author}{\bibfnamefont{A.}~\bibnamefont{Barra}},
  \bibinfo{author}{\bibfnamefont{A.}~\bibnamefont{Galluzzi}},
  \bibinfo{author}{\bibfnamefont{F.}~\bibnamefont{Guerra}},
  \bibinfo{author}{\bibfnamefont{D.}~\bibnamefont{Tantari}}, \bibnamefont{and}
  \bibinfo{author}{\bibfnamefont{F.}~\bibnamefont{Tavani}},
  \bibinfo{journal}{Phys. Rev. E} \textbf{\bibinfo{volume}{91}},
  \bibinfo{pages}{062807} (\bibinfo{year}{2015}{\natexlab{b}}).

\bibitem[{\citenamefont{Agliari
  et~al.}(2015{\natexlab{c}})\citenamefont{Agliari, Barra, Galluzzi, Guerra,
  Tantari, and Tavani}}]{agliari2015hierarchical}
\bibinfo{author}{\bibfnamefont{E.}~\bibnamefont{Agliari}},
  \bibinfo{author}{\bibfnamefont{A.}~\bibnamefont{Barra}},
  \bibinfo{author}{\bibfnamefont{A.}~\bibnamefont{Galluzzi}},
  \bibinfo{author}{\bibfnamefont{F.}~\bibnamefont{Guerra}},
  \bibinfo{author}{\bibfnamefont{D.}~\bibnamefont{Tantari}}, \bibnamefont{and}
  \bibinfo{author}{\bibfnamefont{F.}~\bibnamefont{Tavani}},
  \bibinfo{journal}{Neural Networks} \textbf{\bibinfo{volume}{66}},
  \bibinfo{pages}{22} (\bibinfo{year}{2015}{\natexlab{c}}).

\bibitem[{\citenamefont{Biggs}(1974)}]{Biggs74}
\bibinfo{author}{\bibfnamefont{N.}~\bibnamefont{Biggs}},
  \emph{\bibinfo{title}{Algebraic Graph Theory}} (\bibinfo{publisher}{Cambridge
  University Press}, \bibinfo{year}{1974}).

\bibitem[{\citenamefont{Barahona and Pecora}(2002)}]{Barahona02}
\bibinfo{author}{\bibfnamefont{M.}~\bibnamefont{Barahona}} \bibnamefont{and}
  \bibinfo{author}{\bibfnamefont{L.~M.} \bibnamefont{Pecora}},
  \bibinfo{journal}{Phys. Rev. Lett.} \textbf{\bibinfo{volume}{89}},
  \bibinfo{pages}{054101} (\bibinfo{year}{2002}).

\bibitem[{\citenamefont{Korniss et~al.}(2003)\citenamefont{Korniss, Novotny,
  Guclu, Toroczkai, and Rikvold}}]{Korniss03}
\bibinfo{author}{\bibfnamefont{G.}~\bibnamefont{Korniss}},
  \bibinfo{author}{\bibfnamefont{M.~A.} \bibnamefont{Novotny}},
  \bibinfo{author}{\bibfnamefont{H.}~\bibnamefont{Guclu}},
  \bibinfo{author}{\bibfnamefont{Z.}~\bibnamefont{Toroczkai}},
  \bibnamefont{and} \bibinfo{author}{\bibfnamefont{P.~A.}
  \bibnamefont{Rikvold}}, \bibinfo{journal}{Science}
  \textbf{\bibinfo{volume}{299}}, \bibinfo{pages}{677} (\bibinfo{year}{2003}).

\bibitem[{\citenamefont{Boettcher}(2012)}]{SITIS12}
\bibinfo{author}{\bibfnamefont{S.}~\bibnamefont{Boettcher}}, in
  \emph{\bibinfo{booktitle}{The 9th International Conference on Signal Image
  Technology \& Internet Based Systems (SITIS)}} (\bibinfo{year}{2012}).

\bibitem[{\citenamefont{Maas}(1987)}]{maas87permeability}
\bibinfo{author}{\bibfnamefont{C.}~\bibnamefont{Maas}},
  \bibinfo{journal}{Discrete Applied Mathematics}
  \textbf{\bibinfo{volume}{16}}, \bibinfo{pages}{31} (\bibinfo{year}{1987}).

\bibitem[{\citenamefont{{Boettcher} et~al.}(2011)\citenamefont{{Boettcher},
  {Varghese}, and {Novotny}}}]{Boettcher11b}
\bibinfo{author}{\bibfnamefont{S.}~\bibnamefont{{Boettcher}}},
  \bibinfo{author}{\bibfnamefont{C.}~\bibnamefont{{Varghese}}},
  \bibnamefont{and} \bibinfo{author}{\bibfnamefont{M.~A.}
  \bibnamefont{{Novotny}}}, \bibinfo{journal}{Phys. Rev. E}
  \textbf{\bibinfo{volume}{83}}, \bibinfo{pages}{041106}
  (\bibinfo{year}{2011}).

\bibitem[{\citenamefont{Childs and Goldstone}(2004)}]{Childs04}
\bibinfo{author}{\bibfnamefont{A.~M.} \bibnamefont{Childs}} \bibnamefont{and}
  \bibinfo{author}{\bibfnamefont{J.}~\bibnamefont{Goldstone}},
  \bibinfo{journal}{Phys. Rev. A} \textbf{\bibinfo{volume}{70}},
  \bibinfo{pages}{022314} (\bibinfo{year}{2004}).

\bibitem[{\citenamefont{Agliari et~al.}(2010)\citenamefont{Agliari, Blumen, and
  M\"ulken}}]{PhysRevA.82.012305}
\bibinfo{author}{\bibfnamefont{E.}~\bibnamefont{Agliari}},
  \bibinfo{author}{\bibfnamefont{A.}~\bibnamefont{Blumen}}, \bibnamefont{and}
  \bibinfo{author}{\bibfnamefont{O.}~\bibnamefont{M\"ulken}},
  \bibinfo{journal}{Phys. Rev. A} \textbf{\bibinfo{volume}{82}},
  \bibinfo{pages}{012305} (\bibinfo{year}{2010}).

\bibitem[{\citenamefont{Pothen et~al.}(1990)\citenamefont{Pothen, Simon, and
  Liou}}]{pothen90partitioning}
\bibinfo{author}{\bibfnamefont{A.}~\bibnamefont{Pothen}},
  \bibinfo{author}{\bibfnamefont{H.~D.} \bibnamefont{Simon}}, \bibnamefont{and}
  \bibinfo{author}{\bibfnamefont{K.-P.} \bibnamefont{Liou}},
  \bibinfo{journal}{SIAM journal on matrix analysis and applications}
  \textbf{\bibinfo{volume}{11}}, \bibinfo{pages}{430} (\bibinfo{year}{1990}).

\bibitem[{\citenamefont{Hendrickson and Leland}(1995)}]{HL}
\bibinfo{author}{\bibfnamefont{B.~A.} \bibnamefont{Hendrickson}}
  \bibnamefont{and} \bibinfo{author}{\bibfnamefont{R.}~\bibnamefont{Leland}},
  in \emph{\bibinfo{booktitle}{Proceedings of Supercomputing '95}}
  (\bibinfo{year}{1995}).

\bibitem[{\citenamefont{Peinecke et~al.}(2007)\citenamefont{Peinecke, Wolter,
  and Reuter}}]{peinecke07image}
\bibinfo{author}{\bibfnamefont{N.}~\bibnamefont{Peinecke}},
  \bibinfo{author}{\bibfnamefont{F.-E.} \bibnamefont{Wolter}},
  \bibnamefont{and} \bibinfo{author}{\bibfnamefont{M.}~\bibnamefont{Reuter}},
  \bibinfo{journal}{Computer-Aided Design} \textbf{\bibinfo{volume}{39}},
  \bibinfo{pages}{460} (\bibinfo{year}{2007}).

\bibitem[{\citenamefont{Fisher}(1966)}]{fisher66shape}
\bibinfo{author}{\bibfnamefont{M.~E.} \bibnamefont{Fisher}},
  \bibinfo{journal}{Journal of combinatorial theory}
  \textbf{\bibinfo{volume}{1}}, \bibinfo{pages}{105} (\bibinfo{year}{1966}).

\bibitem[{\citenamefont{Domany et~al.}(1983)\citenamefont{Domany, Alexander,
  Bensimon, and Kadanoff}}]{PhysRevB.28.3110}
\bibinfo{author}{\bibfnamefont{E.}~\bibnamefont{Domany}},
  \bibinfo{author}{\bibfnamefont{S.}~\bibnamefont{Alexander}},
  \bibinfo{author}{\bibfnamefont{D.}~\bibnamefont{Bensimon}}, \bibnamefont{and}
  \bibinfo{author}{\bibfnamefont{L.~P.} \bibnamefont{Kadanoff}},
  \bibinfo{journal}{Phys. Rev. B} \textbf{\bibinfo{volume}{28}},
  \bibinfo{pages}{3110} (\bibinfo{year}{1983}).

\bibitem[{\citenamefont{Rammal}(1984)}]{Rammal84}
\bibinfo{author}{\bibfnamefont{R.}~\bibnamefont{Rammal}}, \bibinfo{journal}{J.
  Physique} \textbf{\bibinfo{volume}{45}}, \bibinfo{pages}{191}
  (\bibinfo{year}{1984}).

\bibitem[{\citenamefont{Fukushima and Shima}(1992)}]{fukushima92spectral}
\bibinfo{author}{\bibfnamefont{M.}~\bibnamefont{Fukushima}} \bibnamefont{and}
  \bibinfo{author}{\bibfnamefont{T.}~\bibnamefont{Shima}},
  \bibinfo{journal}{Potential Analysis} \textbf{\bibinfo{volume}{1}},
  \bibinfo{pages}{1} (\bibinfo{year}{1992}).

\bibitem[{\citenamefont{Rammal and Toulouse}(1983)}]{rammal83random}
\bibinfo{author}{\bibfnamefont{R.}~\bibnamefont{Rammal}} \bibnamefont{and}
  \bibinfo{author}{\bibfnamefont{G.}~\bibnamefont{Toulouse}},
  \bibinfo{journal}{Journal de Physique Lettres} \textbf{\bibinfo{volume}{44}},
  \bibinfo{pages}{13} (\bibinfo{year}{1983}).

\bibitem[{\citenamefont{Shima}(1991)}]{shima91eigenvalue}
\bibinfo{author}{\bibfnamefont{T.}~\bibnamefont{Shima}},
  \bibinfo{journal}{Japan Journal of Industrial and Applied Mathematics}
  \textbf{\bibinfo{volume}{8}}, \bibinfo{pages}{127} (\bibinfo{year}{1991}).

\bibitem[{\citenamefont{Teplyaev}(2000)}]{teplyaev00gradients}
\bibinfo{author}{\bibfnamefont{A.}~\bibnamefont{Teplyaev}},
  \bibinfo{journal}{Journal of Functional Analysis}
  \textbf{\bibinfo{volume}{174}}, \bibinfo{pages}{128} (\bibinfo{year}{2000}).

\bibitem[{\citenamefont{Lyons}(2005)}]{Lyons05}
\bibinfo{author}{\bibfnamefont{R.}~\bibnamefont{Lyons}},
  \bibinfo{journal}{Combin. Probab. Comput.} \textbf{\bibinfo{volume}{14}},
  \bibinfo{pages}{491} (\bibinfo{year}{2005}).

\bibitem[{\citenamefont{Dhar}(1999)}]{Dhar99}
\bibinfo{author}{\bibfnamefont{D.}~\bibnamefont{Dhar}},
  \bibinfo{journal}{Physica A} \textbf{\bibinfo{volume}{263}},
  \bibinfo{pages}{4} (\bibinfo{year}{1999}).

\bibitem[{\citenamefont{Wu et~al.}(2006)\citenamefont{Wu, Braunstein, Havlin,
  and Stanley}}]{wutransport:06}
\bibinfo{author}{\bibfnamefont{Z.}~\bibnamefont{Wu}},
  \bibinfo{author}{\bibfnamefont{L.~A.} \bibnamefont{Braunstein}},
  \bibinfo{author}{\bibfnamefont{S.}~\bibnamefont{Havlin}}, \bibnamefont{and}
  \bibinfo{author}{\bibfnamefont{H.~E.} \bibnamefont{Stanley}},
  \bibinfo{journal}{Phys. Rev. Lett.} \textbf{\bibinfo{volume}{96}},
  \bibinfo{pages}{148702} (\bibinfo{year}{2006}).

\bibitem[{\citenamefont{Nishikawa and Motter}(2006)}]{Nishikawa:06}
\bibinfo{author}{\bibfnamefont{T.}~\bibnamefont{Nishikawa}} \bibnamefont{and}
  \bibinfo{author}{\bibfnamefont{A.~E.} \bibnamefont{Motter}},
  \bibinfo{journal}{Phys. Rev. E} \textbf{\bibinfo{volume}{73}},
  \bibinfo{pages}{065106} (\bibinfo{year}{2006}).

\bibitem[{\citenamefont{Dhar}(1977)}]{Dhar77}
\bibinfo{author}{\bibfnamefont{D.}~\bibnamefont{Dhar}}, \bibinfo{journal}{J.
  Math. Phys.} \textbf{\bibinfo{volume}{18}}, \bibinfo{pages}{578}
  (\bibinfo{year}{1977}).

\bibitem[{\citenamefont{Wu}(1982)}]{Wu82}
\bibinfo{author}{\bibfnamefont{F.~Y.} \bibnamefont{Wu}},
  \bibinfo{journal}{Reviews of Modern Physics} \textbf{\bibinfo{volume}{54}},
  \bibinfo{pages}{235} (\bibinfo{year}{1982}).

\bibitem[{\citenamefont{Shrock and Wu}(2000)}]{shrock:00}
\bibinfo{author}{\bibfnamefont{R.}~\bibnamefont{Shrock}} \bibnamefont{and}
  \bibinfo{author}{\bibfnamefont{F.~Y.} \bibnamefont{Wu}},
  \bibinfo{journal}{Journal of Physics A: Mathematical and General}
  \textbf{\bibinfo{volume}{33}}, \bibinfo{pages}{3881} (\bibinfo{year}{2000}).

\bibitem[{\citenamefont{Chang et~al.}(2007)\citenamefont{Chang, Chen, and
  Yang}}]{Chang07}
\bibinfo{author}{\bibfnamefont{S.-C.} \bibnamefont{Chang}},
  \bibinfo{author}{\bibfnamefont{L.-C.} \bibnamefont{Chen}}, \bibnamefont{and}
  \bibinfo{author}{\bibfnamefont{W.-S.} \bibnamefont{Yang}},
  \bibinfo{journal}{J. Stat. Phys.} \textbf{\bibinfo{volume}{126}},
  \bibinfo{pages}{649} (\bibinfo{year}{2007}).

\bibitem[{\citenamefont{Teufl and Wagner}(2006)}]{teufl06spanning}
\bibinfo{author}{\bibfnamefont{E.}~\bibnamefont{Teufl}} \bibnamefont{and}
  \bibinfo{author}{\bibfnamefont{S.}~\bibnamefont{Wagner}},
  \bibinfo{journal}{DMTCS Proceedings}  (\bibinfo{year}{2006}).

\bibitem[{\citenamefont{Teufl and Wagner}(2011)}]{teufl11Strees}
\bibinfo{author}{\bibfnamefont{E.}~\bibnamefont{Teufl}} \bibnamefont{and}
  \bibinfo{author}{\bibfnamefont{S.}~\bibnamefont{Wagner}},
  \bibinfo{journal}{Journal of Statistical Physics}
  \textbf{\bibinfo{volume}{142}}, \bibinfo{pages}{879} (\bibinfo{year}{2011}).

\bibitem[{\citenamefont{Teufl and Wagner}(2010)}]{Teufl10}
\bibinfo{author}{\bibfnamefont{E.}~\bibnamefont{Teufl}} \bibnamefont{and}
  \bibinfo{author}{\bibfnamefont{S.}~\bibnamefont{Wagner}},
  \bibinfo{journal}{J. Phys. A: Math. Theor.} \textbf{\bibinfo{volume}{43}},
  \bibinfo{pages}{415001} (\bibinfo{year}{2010}).

\bibitem[{Note1()}]{Note1}
 \bibinfo{note}{The hierarchy index $i$ of sites $n$ in Eq. (\ref
  {eq:numbering}) resembles the sequence by which discs are moved in the famous
  ``Tower of Hanoi'' problem. Unfortunately, there also exists a ``Hanoi
  graph'' (see Weisstein, Eric W. \textquotedbl {}Hanoi Graph.\textquotedbl {}
  From MathWorld--A Wolfram Web Resource.
  http://mathworld.wolfram.com/HanoiGraph.html), essentially the dual of the
  Sierpinski gasket, that should not be confused with our networks here.}

\bibitem[{\citenamefont{Ramond}(1997)}]{Ramon97}
\bibinfo{author}{\bibfnamefont{P.}~\bibnamefont{Ramond}},
  \emph{\bibinfo{title}{Field Theory: A Modern Primer}}
  (\bibinfo{publisher}{Westview Press}, \bibinfo{year}{1997}).

\bibitem[{\citenamefont{Hasegawa and Nogawa}({2013})}]{Hasegawa13c}
\bibinfo{author}{\bibfnamefont{T.}~\bibnamefont{Hasegawa}} \bibnamefont{and}
  \bibinfo{author}{\bibfnamefont{T.}~\bibnamefont{Nogawa}},
  \bibinfo{journal}{Phys. Rev. E} \textbf{\bibinfo{volume}{{87}}},
  \bibinfo{pages}{032810} (\bibinfo{year}{{2013}}).

\bibitem[{\citenamefont{Abramowitz and Stegun}(1964)}]{abramowitz:64}
\bibinfo{author}{\bibfnamefont{M.}~\bibnamefont{Abramowitz}} \bibnamefont{and}
  \bibinfo{author}{\bibfnamefont{I.~A.} \bibnamefont{Stegun}},
  \emph{\bibinfo{title}{{Handbook of Mathematical Functions with Formulas,
  Graphs, and Mathematical Tables}}} (\bibinfo{publisher}{Dover},
  \bibinfo{address}{New York}, \bibinfo{year}{1964}).

\bibitem[{\citenamefont{Boettcher
  et~al.}(2008{\natexlab{b}})\citenamefont{Boettcher, Gon{\c{c}}alves, and
  Azaret}}]{SWlong}
\bibinfo{author}{\bibfnamefont{S.}~\bibnamefont{Boettcher}},
  \bibinfo{author}{\bibfnamefont{B.}~\bibnamefont{Gon{\c{c}}alves}},
  \bibnamefont{and} \bibinfo{author}{\bibfnamefont{J.}~\bibnamefont{Azaret}},
  \bibinfo{journal}{J. Phys. A: Math. Theor.} \textbf{\bibinfo{volume}{41}},
  \bibinfo{pages}{335003} (\bibinfo{year}{2008}{\natexlab{b}}).

\bibitem[{\citenamefont{Livio}(2003)}]{Livio03}
\bibinfo{author}{\bibfnamefont{M.}~\bibnamefont{Livio}},
  \emph{\bibinfo{title}{The Golden Ratio: The Story of $\Phi$, the World's Most
  Astonishing Number}} (\bibinfo{publisher}{Broadway Books},
  \bibinfo{address}{New York}, \bibinfo{year}{2003}).

\end{thebibliography}

\end{document}